%% file: paper.tex
\documentclass{llncs}

\title{Starting a Dialog between Model Checking and Fault-tolerant
     Distributed Algorithms\thanks{Supported in part by the Austrian National Research
    Network S11403-N23 (RiSE) of the Austrian Science Fund (FWF),
    and by the Vienna Science and Technology Fund (WWTF)
    grant PROSEED.}
}

\author{Annu John \and Igor Konnov \and Ulrich Schmid
    \and Helmut Veith \and Josef Widder}
\institute{Vienna University of Technology (TU Wien)}

\usepackage{lscape}

\usepackage{longtable}

\usepackage{hyperref}
\usepackage{listings}
\usepackage{subfig}
\usepackage{stmaryrd}
\date{}

\usepackage{color, colortbl}

\usepackage{pifont}
\newcommand{\xmark}{\ding{55}}
\newcommand{\checkmark}{\ding{51}}

 \newcommand{\exclude}[1]{
 }
\usepackage{amsfonts,amssymb,amsmath}

\lstdefinelanguage{promela}
  {morekeywords={do,od,init,proctype,if,fi,byte,bool,atomic},
  morecomment=[s]{/*}{*/},escapechar=\@,
  basicstyle=\small\ttfamily,
  commentstyle=\itshape\rmfamily\small,
  keywordstyle=\ttfamily\small\underbar
}

\usepackage{tikz}
\usetikzlibrary{arrows}
\usetikzlibrary{shapes}
\usetikzlibrary{calc}
\usetikzlibrary{positioning}

\newcommand{\ident}[1]{\textit{#1\/}\rule{0cm}{1ex}}
\gdef\dash---{\thinspace---\hskip.16667em\relax}
\gdef\ndash---{\thinspace--\hskip.16667em\relax}

\newcommand{\Natural}{{\mathbb N}}

\newcommand{\trmove}{{\sc (move)}}
\newcommand{\trmaintain}{{\sc (frame)}}

\newcommand{\paraset}{\Pi}
\newcommand{\globset}{\Gamma}
\newcommand{\locset}{\Lambda}
\newcommand{\pcval}{Z}

\newcommand{\AdmP}{\mathbf{P}_{RC}}
\newcommand{\param}{\mathbf{p}}

\newcommand{\pc}{\ident{sv}}
\newcommand{\PC}{\ident{SV}}
\newcommand{\rcvd}{\ident{rcvd}}
\newcommand{\sent}{\ident{nsnt}}
\newcommand{\numparam}{{|\paraset|}}

\newcommand{\restrict}[2]{#1 |_{#2}}

\newcommand{\syssize}{N}

\newcommand{\true}{\text{\sc true}}
\newcommand{\false}{\text{\sc false}}

\newcommand{\Prop}{\text{AP}}
\newcommand{\PropPC}{\Prop_{\PC}}
\newcommand{\PropVAR}{\Prop_\ident{D}}
\newcommand{\aprop}{\ident{p}}

\newcommand{\gleich}[1]{=_{#1}}

\newcommand{\ResCond}{{\ident{RC}}} 

\newcommand{\CFA}{A}

\newcommand{\Sk}{\textsf{Sk}}
\newcommand{\SkAut}{\textsf{Sk}(\CFA)}

\newcommand{\instBLA}{\textsf{Inst}}

\newcommand{\sConcSys}{{\instBLA(\param,\Sk)}}
\newcommand{\ConcSys}{{\instBLA(\param,\SkAut)}}

\newcommand{\LetterConcSys}{\textsf{Inst}}

\newcommand{\newreftheorem}[2]{
 \newenvironment{#1}[1]{\par\vspace{3mm}\noindent\textbf{#2~\ref{##1}.}
\em}{\rm}
}
\newreftheorem{reflemma}{Lemma}
\newreftheorem{refproposition}{Proposition}
\newreftheorem{reftheorem}{Theorem}
\newreftheorem{refcorollary}{Corollary}

\newcommand{\CFAtemplate}[1]{{\mathtt{#1}}}
\newcommand{\CFAreserved}[1]{{\underline{\mathtt{#1}}}}
\newcommand{\CFAloc}{\CFAtemplate{sval}}
\newcommand{\CFAvar}{\CFAtemplate{var}}
\newcommand{\CFAcvar}{\CFAtemplate{condvar}}
\newcommand{\CFAparam}{\CFAtemplate{param}}
\newcommand{\CFAparamcombi}{\CFAtemplate{lin\_form}}
\newcommand{\CFAthreshold}{\CFAtemplate{threshold}}
\newcommand{\CFAguard }{\CFAtemplate{guard}}
\newcommand{\CFAcond}{\CFAtemplate{cond}}
\newcommand{\CFAOp}{\CFAtemplate{Op}}
\newcommand{\CFApc}{{\pc}}
\newcommand{\CFAinc}{\CFAreserved{inc}}

\newcommand{\CFAcondatom}{\CFAtemplate{atomcond}}
\newcommand{\CFAdummy}{\varepsilon}
\newcommand{\CFApick}{\; \CFAreserved{where} \;}
\newcommand{\CFApickOp}[3]{#1 #2 #3}
\newcommand{\sem}[1]{\llbracket #1 \rrbracket}

\newcommand\labelfun{\lambda}

\newcommand\gst{\sigma}

\newcommand{\IT}{\mathrm{V0}}
\newcommand{\RI}{\mathrm{V1}}
\newcommand{\SE}{\mathrm{SE}}
\newcommand{\AC}{\mathrm{AC}}

\newcommand{\accept}{\text{accept}}
\newcommand{\echomsg}{\langle\text{echo}\rangle}

\newcommand\ltlF{\textsf{\textbf{F}}\,}
\newcommand\ltlG{\textsf{\textbf{G}}\,}
\newcommand\ltlU{\,\textsf{\textbf{U}}\,}
\newcommand\LTLX{$\mbox{\textsf{LTL}} \setminus \textsf{{X}}$}

\newcommand\Wedge[2]{\ensuremath{\bigwedge\limits_{#1}^{#2}}}
\newcommand\Vee[2]{\ensuremath{\bigvee\limits_{#1}^{#2}}}

\usepackage{color}

\usepackage{ifthen}

\usepackage{algorithm}
\usepackage[noend]{algorithmic}

\makeatletter
\newcommand{\EMPTY}{\item[]}

\newcommand{\newlinetag}[3]{\newcommand{#1}[#2]{\item[#3]}}
\newcommand{\newconstruct}[5]{
  \newenvironment{ALC@\string#1}{\begin{ALC@g}}{\end{ALC@g}}
   \newcommand{#1}[2][default]{\ALC@it#2\ ##2\ #3
     \ALC@com{##1}\begin{ALC@\string#1}}
   \ifthenelse{\boolean{ALC@noend}}{
     \newcommand{#4}{\end{ALC@\string#1}}
   }{
     \newcommand{#4}{\end{ALC@\string#1}\ALC@it#5}
   }
}

\newcommand{\ALCEXT@linenosize}{\small}
\newcommand{\ALCEXT@linenofont}{\rm}
\newcommand{\renew@ALC@linenosize}{\renewcommand{\ALC@linenosize}{\ALCEXT@linenosize\ALCEXT@linenofont}}
\newcommand{\setlinenosize}[1]{\renewcommand{\ALCEXT@linenosize}{#1}\renew@ALC@linenosize}
\newcommand{\setlinenofont}[1]{\renewcommand{\ALCEXT@linenofont}{#1}\renew@ALC@linenosize}

\renew@ALC@linenosize
\renewcommand{\ALC@linenodelimiter}{:}

\newcounter{ALCEXT@lineno}

\let\ALCEXT@endalgorithmic=\endalgorithmic
\def\endalgorithmic{\setcounter{ALCEXT@lineno}{\value{ALC@line}}\ALCEXT@endalgorithmic}

\newlinetag{\CODE}{1}{\textrm{Code for processes #1:}}

\newlinetag{\PARAM}{0}{\textbf{Parameters}}
\newlinetag{\VAR}{0}{\textbf{Variables}}
\newlinetag{\TRANS}{0}{\textbf{Rules}}
\makeatother

\makeatletter
\newcommand{\MESGitem}[1]{\ensuremath{#1}}
\newcommand{\MESGsep}{, }

\def\MESG#1{{(\/\let\theMessage={}\parsefirst#1,\@end,\theMessage\/)}}
\def\parsefirst#1,{
  \ifx#1\@end \let\next=\relax\else
  \let\next=\parsemessage
  \let\theMessage={\theMessage\MESGitem{#1}}\fi \next}
\def\parsemessage#1,{
  \ifx#1\@end \let\next=\relax \else
  \let\next=\parsemessage
  \let\theMessage={\theMessage\MESGsep\MESGitem{#1}}\fi \next}

\def\new@MESG#1[#2]{\expandafter\gdef\csname #1\endcsname##1{
    \MESG{\textsc{#2},##1}}}

\def\newMESG#1{\@ifnextchar [{\new@MESG{#1}}{\new@MESG{#1}[#1]}}
\makeatother

\newcommand{\codelineref}[1]{line~{\small\tt\ref{#1}}}
\newcommand{\codelinerange}[2]{lines~{\small\tt\ref{#1}-\ref{#2}}}

\begin{document}

\pagestyle{plain}

\maketitle

\begin{abstract}
Fault-tolerant distributed algorithms are central for building
     reliable spatially distributed systems.
Unfortunately, the lack of a canonical precise framework for
     fault-tolerant algorithms is an obstacle for both verification
     and deployment.
In this paper, we introduce a new domain-specific framework to capture
     the behavior of fault-tolerant distributed algorithms in an
     adequate and precise way.
At the center of our framework is a parameterized system model where
     control flow automata are used for process specification.
To account for the specific features and properties of fault-tolerant
     distributed algorithms for message-passing systems,  our control
     flow automata are extended to model threshold guards as well as
     the inherent non-determinism stemming from asynchronous
     communication,  interleavings of steps, and faulty processes.

We demonstrate the adequacy of our framework in a representative case
     study where we formalize a family of well-known   fault-tolerant
     broadcasting algorithms under a variety of failure assumptions.
Our case study is supported by model checking experiments with safety
     and liveness specifications for a fixed number of processes.
In the experiments, we systematically varied the assumptions on both
     the resilience condition and the failure model.
In all cases, our experiments coincided with the theoretical results
     predicted in the distributed algorithms literature.
This is giving clear evidence for the adequacy of our model.

In a companion paper~\cite{JKSVW12a}, we are addressing the new model
     checking techniques necessary for parametric verification of the
     distributed algorithms captured in our framework.
\end{abstract}

\section{Introduction}

Even formally verified computer systems are subject to power outages,
     bad electrical connections, arbitrary bit-flips in memory, etc.
A classic approach to ensure that a computer system is reliable, and
     continues to perform its task, even if some components fail, is
     replication.
The idea is to have multiple computers instead of a single one (that
     would constitute a single point of failure), and ensure that the
     replicated computers coordinate, and for instance in the case of
     replicated databases, store the same information.
Ensuring that all computers agree on the same information is
     non-trivial due to several sources of non-determinism, namely,
     faults, uncertain message delays, and asynchronous
     computation steps.

To address this important problem, fault-tolerant distributed
     algorithms were introduced many years ago.
As they are designed to increase the reliability of systems, it is
     crucial that they are in fact correct, i.e., that they satisfy their
     specifications.
Due to the mentioned non-determinism it is however very easy to make
     mistakes in the correctness arguments for fault-tolerant
     distributed algorithms.
The combination of criticality and difficulty make fault-tolerant
distributed algorithms a natural candidate for model checking.

Unfortunately, there are three reasons why model checking
     fault-tolerant distributed algorithms is a complicated task.
(i) First, they are mathematically complex, and inherently contain
     many sources of non-determinism.
(ii) Second, there is no canonical model, such that each algorithm
     comes with different\dash---and usually
     complex\dash---assumptions about the environment, in particular
     assumptions on degrees of concurrency, message delays, and
     failure models.
By failure models we mean both, the anticipated behavior of faulty
     components and a resilience condition stating the fraction of
     faulty components among all components of the system.
(iii) Finally, distributed algorithms are traditionally described in
     pseudocode.
This approach is problematic because every paper comes with a
     different (alas unspecified) pseudo code language.
It is often not clear how a given pseudo code is related to the
     computational model that is provided only in natural language.
Other authors from formal methods have also argued that the algorithms
     and proofs in these papers are hard to understand for
     outsiders~\cite{FuzzatiMN07}.

All these problems result in three verification research problems for
     the use of fault-tolerant distributed algorithms.
Together, they close the methodological gap between distributed
     algorithms, verification, and software engineering: 

\begin{description}
\item[Formalization problem.] There is no modeling language for fault tolerant distributed algorithms, and the
     pseudocode and hidden assumptions make it difficult to understand
     the semantics.
Success in this area crucially depends on collaboration of researchers
     from model checking and from distributed algorithms.

\item[Verification problem.] Even in the presence of a precise model, there
     are many open problems in the area of verification.
A central challenge is  parameterized model checking, i.e., verification of
     a given fault-tolerant distributed algorithm for all system
     sizes.
Note however that verification without adequate formalization is pointless, as
     one can never be sure what actually has been verified.
\item[Deployment problem.]
How can one transfer a formal model to a real-life implementation and ensure
     its conformance with the (verified) model?
\end{description}

This paper is devoted to the formalization problem. In a companion paper~\cite{JKSVW12a}, we
are developing new parameterized model checking methods for fault tolerant distributed algorithms.
A central and important goal of our work is to initiate a systematic
     study of distributed algorithms from a verification and
     programming language point of view in a way that does not betray
     the fundamentals of distributed algorithms.
The famous bakery algorithm~\cite{Lamport74a} is the most striking example from
the literature where wrong specifications have been verified or wrong semantics
have been considered. Many papers in formal methods have claimed to prove
correctness of the bakery algorithm as evidence for their practical applicability. From a distributed
algorithms perspective, however,
 most of these papers miss that the
     algorithm is \emph{not} using the strong
     assumption of atomic registers but requires only safe
     registers~\cite{Lamport86}.
Although this example shows that many issues in distributed algorithms
are quite subtle, the distributed algorithm literature is often not very explicit about
them, making it hard for non-experts to extract the correct model.

Reviewing the distributed algorithms literature, the formalization
problem can be reduced to two central questions. (i) First, the question
how algorithms are described, and (ii) second, the question how to capture
the vast diversity in computational models that describe the environment.

\noindent{(i)} Algorithm descriptions in the literature are
     based on pseudo code, whose semantics is described in a
     handwaving manner (if it is described at all): In particular,
details which  are considered not interesting for the current
     distributed algorithm are ignored or only hinted at, e.g., the
     bookkeeping over the messages that have been sent and received so far.

\noindent{(ii)} The example of the bakery algorithm shows that
     assumptions on computational models are very subtle.
The distributed algorithms literature does, however, rarely state
     these assumptions precisely but rather present them in natural
     language.
This is very unfortunate as there are quite involved assumptions that
     are usually not considered in model checking.
For instance, \cite{DLS88} postulates that in each run there is a
     value $\Phi$ such that between two steps of a process, every
     other process takes at most $\Phi$ steps.
Assumptions of this kind are crucial for fault-tolerant distributed
     algorithms because there are impossibility results for the
     classic asynchronous interleaving semantics~\cite{FLP85}.
In other words, without these assumptions, all algorithms violate the
     specification.
Apart from interleaving of steps, non-trivial assumptions can also be
     found for message delays, behavior of faulty processes, behavior
     of faulty links, and resilience conditions on the fraction of
     faulty process.

To address both aspects of the formalization problem, we need a novel framework which
is natural and adequate for the distributed algorithms community, but precise
enough to facilitate automated verification. The current paper represents a
first step towards this goal.

\paragraph{Contributions of the current paper.}

We introduce a new framework for the specification of distributed
     algorithms.
We focus on the important class of threshold-guarded fault-tolerant
     distributed algorithms, which we discuss in detail in
     Section~\ref{sec:Intro}.
For this class, we introduce a parameterized modeling framework based
     on control flow automata, which is a notion from software model
     checking extended by non-determinism  and threshold guards.
Our framework facilitates flexible fine-grained and adequate
     description of distributed algorithms under different fault
     assumptions and resilience conditions.
For automatic treatment, we have a front-end similar to
     Promela~\cite{H2003}.
With this formalism we can express, e.g., several variants of classic
     asynchronous broadcasting algorithms~\cite{ST87:abc} under
     different fault assumptions.
Our framework is only the first step towards considering various
     environments that are different from the asynchronous one, e.g.,
     partial synchrony or round models.
This will allow us to express a wider range of fault-tolerant
     distributed algorithms, e.g.,
     \cite{DLS88,ChandraT96,Charron-BostS09,FS12:DC,WS07:DC}.

After introducing the framework, we provide a case study and
     experiments that show how to translate distributed algorithms into
     our framework.
We discuss the formalization of an important family of fault-tolerant
     distributed algorithms, which is  well-understood.
In our experiments we consider different fault models, and
     systematically validate the adequacy of our modeling framework.
The experiments are made for a bounded number of processes, i.e., the
     model checking is relatively straight-forward, but not scalable
     to large numbers of processes.
As we do not need abstractions for this tasks, we avoid simplification
     artefacts.
Our experiments show that our modeling framework is adequate.
Thus we have a starting point for a serious investigation of
     parameterized model checking and of systematic deployment.

\paragraph{Organization of the paper.}

We first discuss the standard pseudocode construct of fault-tolerant
     distributed algorithms\dash---namely threshold guards\dash---in
     Section~\ref{sec:Intro}.
In Section~\ref{sec:prelim}, we introduce a general system model for
     fault-tolerant distributed algorithms that provides parameterized
     processes, parameterized system sizes, and resilience conditions.
Section~\ref{sec:DA} introduces our novel variant of control flow
     automata, and discusses how they can be composed to derive
     instances of distributed systems based on the model of
     Section~\ref{sec:prelim}.
In Section~\ref{sec:casestudy}, we describe the translation of our
     case study algorithm to its corresponding control flow automaton.
Section~\ref{sec:exp} presents the outcomes of our model checking
     experiments, and Section~\ref{sec:relwork} relates our approach
     to other existing approaches for specifying concurrent
     algorithms.


\section{Threshold guarded distributed algorithms}\label{sec:Intro}

Processes that execute the instances of a distributed algorithm
     exchange messages, and the state transitions of these processes
     are predominantly determined by the messages received.
In addition to the standard execution of actions, which are guarded by
     some predicate on the local state, most basic distributed
     algorithms (cf.~\cite{Lyn96,AW04}) just add existentially and/or
     universally guarded commands involving received messages:
\begin{center}
\begin{minipage}{.45\linewidth}
  \begin{lstlisting}[language=pascal,label=Lst:ExistsGC]
    if received <m>
       from some process
    then action(m);
  \end{lstlisting}
\begin{center}
       (a) existential guard
\end{center}
\end{minipage}
\begin{minipage}{.45\linewidth}
  \begin{lstlisting}[language=pascal,label=Lst:ForallCG]
   if received <m>
      from all processes
   then action(m);
  \end{lstlisting}
\begin{center}
       (b) universal guard
\end{center}
\end{minipage}
\end{center}

\smallskip

Depending on the content of the message~{\tt <m>}, the function {\tt
     action} performs a local state transition and possibly sends
     messages to one or more processes.
Such constructs can be found, e.g., in (non-fault-tolerant)
     distributed algorithms for constructing spanning trees, flooding,
     or network synchronization~\cite{Lyn96}.

Understanding and analyzing such distributed algorithms is far from being
trivial, which is due to the uncertainty that local processes have about
the state of other processes. After all, real processors execute at
different and varying speeds, and the end-to-end message delays
also vary considerably. Viewed from the global perspective, this
results in considerable non-determinism of the executions of a distributed
system.

Another very important additional source of non-determinism are faults.
In fact, one of the major benefits of using distributed algorithms is their ability
to cope with faults. In case of distributed agreement,
for example, it is guaranteed that all non-faulty processes compute
the same result even if some other processes fail. Fault-tolerant
distributed algorithms hence typically increase the reliability of
distributed systems \cite{Pow92}.

In order to shed some light on the difficulties faced by a distributed
     algorithm in the presence of faults, consider Byzantine faults
     \cite{LSP80}, which allow a faulty process to behave arbitrarily:
     Faulty processes may fail to send messages, send messages with
     erroneous values, or even send conflicting information to
     different processes.
In addition, faulty processes may even collaborate in order to
     increase their adverse power.
In practice, Byzantine faults can be caused by power outages, bad
     electrical connections, arbitrary bit-flips in memory, or even
     unexpected behavior due to intruders who have taken over control
     of some part of the system.

If one used the construct of Example~(a) in the presence of Byzantine
     faults, the (local state of the) receiver process would be
     corrupted if the received message {\tt <m>} originates in a
     faulty process.
A faulty process could hence contaminate a correct process.
On the other hand, if one tried to use the construct of Example~(b), a
     correct process would wait forever (starve) when a faulty process
     omits to send the required message.
To overcome those problems, fault-tolerant distributed algorithms
     typically require assumptions on the maximum number of faults,
     and employ suitable thresholds for the number of  messages which
     can  be expected to be received by correct processes.
Assuming that the system consists of  $n$ processes among which at
     most $t$ may be faulty,  \emph{threshold guarded commands} such
     as the following are typically used by fault-tolerant distributed
     algorithms:
\begin{center}
\begin{minipage}{.8\linewidth}
  \begin{lstlisting}[language=pascal,label=Lst:ThresGC]
   if received <m> from n-t distinct processes
   then action(m);
  \end{lstlisting}
\end{minipage}
\end{center}
Assuming that thresholds are functions of the
parameters~$n$ and~$t$, threshold guards are a just generalization of
quantified guards as given in Examples~(a) and~(b):
In the above command, a process waits to receive $n-t$ messages from
     distinct processes.
As there are at least $n-t$ correct processes, the guard
cannot be blocked by faulty processes,
     which avoids the problems of the construct of Example~(b).
In the distributed algorithms literature, one finds a variety of
different threshold guarded commands. Another prominent example is $t+1$,
which ensures that at least one message comes from a non-faulty
process.

However, in the setting of Byzantine fault tolerance, it is important
to note that the use of threshold
guarded commands implicitly rests on the assumption that a receiver
can distinguish messages from different senders. In practice,
this can be achieved e.g.\ by using point-to-point links between
processes or by message authentication. What is important here
is that Byzantine faulty processes are only allowed to exercise
control on their own messages and computations, but not on the
messages sent by other processes and the computation of other
processes.

\section{Parameterized System Model}\label{sec:prelim}


We model distributed algorithms via their parameters, the processes,
     and the communication medium, the latter via shared variables.
In Section~\ref{sec:DA}, we will introduce a new variant of control
     flow automata that allows to specify processes of fault-tolerant
     distributed algorithms.
We will discuss how message passing distributed algorithms (as
     mentioned in Section~\ref{sec:Intro}) can be expressed in such a
     model in Section~\ref{sec:casestudy}.

We shall define the parameters, local variables of the processes, and
     shared variables referring to a single \emph{domain} $D$ that is
     totally ordered and has the operations addition and subtraction.
In this paper we will assume that $D = \Natural_0$.
We use the standard notion of models denoted by $\models$.


We start with some notation.
Let $Y$ be a finite set of variables ranging over~$D$. We will denote
     by $D^{|Y|}$, the set of all $|Y|$-tuples of variable values.
In order to simplify notation, given $s\in D^{|Y|}$, we use the
     expression $s.y$, to refer to the value of a variable $y  \in Y$
     in vector $s$.
For two vectors of variable values $s$ and $s'$, by $s
     \gleich{X} s'$ we denote the case where for all $x\in X$, $s.x = s'.x$ holds.

\newcommand{\vars}{V}

The finite set of variables $\vars = \paraset \cup \{\pc\} \cup
     \locset \cup \globset$, where the separate sets are described
     below.
The finite set $\paraset$ is a set of \emph{parameter variables} that
     range over~$D$, and the \emph{resilience condition} $\ResCond$ is
     a predicate over $D^{\numparam}$.
In our example, $\paraset = \{n,t,f\}$, and the resilience condition
     $\ResCond(n,t,f)$ is $n>3t \,\wedge\, f \le t \,\wedge\, t>0$.
Then, we denote the set of \emph{admissible parameters} by  $\AdmP =
     \{ \param\in D^\numparam \mid RC(\param)\}$.
The variable $\pc$ is the  \emph{status variable} that ranges over a
     finite set $\PC$ of \emph{status values}.
(For simplicity, we assume that only one status variable is used;
     however, multiple finite domain status variables can be encoded into
     $\pc$.) 
The finite set $\locset$ contains variables that range over the
     domain~$D$.
The variable $\pc$ and the variables from $\locset$ are \emph{local
     variables}.
The finite set $\globset$ contains the \emph{shared variables} that
     range over $D$.

A process operates on states from the set $S=\PC \times D^{|\locset|}
     \times D^{|\globset|} \times D^{\numparam}$.
Each process starts its computation in an initial state from a set
     $S^0\subseteq S$.
A relation $R \subseteq S \times S$ defines \emph{transitions} from
     one state to another, with the restriction that the  values of
     parameters remain unchanged, i.e., for all  $(s,
     t) \in R$, $s \gleich{\paraset} t$.
Then, a \emph{parameterized process skeleton} is a tuple~$\Sk = (S, S^0,
     R)$.

We get a process instance by fixing the parameter values $\param
     \in D^{\numparam}$: 
one can restrict the set of process states to $\restrict{S}{\param} =
     \{ s \in S \mid s \gleich{\paraset} \param \}$ as well as the set
     of transitions to
     $\restrict{R}{\param} = R \cap (\restrict{S}{\param} \times
     \restrict{S}{\param})$.
Then, a \emph{process instance}  is a process skeleton $\restrict{\Sk}{\param} =
     (\restrict{S}{\param}, \restrict{S^0}{\param},
     \restrict{R}{\param})$ where $\param$ is constant.

For fixed admissible parameters $\param$, a distributed system is
     modeled as an asynchronous parallel composition of identical
     processes $\restrict{\Sk}{\param}$.
The number of processes in this parallel composition depends on the
     parameters.
To formalize this, we define the size of a system (the number of
     processes) using a  function $\syssize\colon \AdmP \rightarrow
     \Natural$.
On our example, we will model only non-faulty processes explicitly in
     our case study, and we will thus use $n - f$ for
     $\syssize(n,t,f)$ in our case~study.

Finally, given $\param \in \AdmP$, and a parameterized process
     skeleton $\Sk = (S, S^0, R)$, we can define a \emph{system
     instance} as a Kripke structure.
Let $\Prop$ be a set of atomic propositions.
(The specific atomic propositions and labeling function that we will
     consider in this paper will be introduced in
     Section~\ref{sec:APs}.) A \emph{system instance} $\sConcSys$ is a
     Kripke structure $(S_\LetterConcSys, S^0_\LetterConcSys,
     R_\LetterConcSys, \Prop, \labelfun_\LetterConcSys)$ where:

\begin{itemize}
\item The set of \emph{(global) states} is $S_\LetterConcSys = \{ (\gst[1], \dots, \gst[\syssize(\param)]) \in (\restrict{S}{\param})^{\syssize(\param)}
\mid \forall i, j \in \{1,
\dots, \syssize(\param) \},  \gst[i] \gleich{\globset \cup  \paraset}
     \gst[j] \}$.
More informally, a global state $\gst$ is a Cartesian product of the
     state $\gst[i]$ of each process~$i$, where the values of
     parameters and shared variables are the same at each process.

\item $S^0_\LetterConcSys = (S^0)^{\syssize(\param)} \cap S_\LetterConcSys$ is the set of
     \emph{initial (global) states}, where $(S^0)^{\syssize(\param)}$
     is the Cartesian product of initial states of individual
     processes.

\item A transition $(\gst, \gst')$ from a global state $\gst \in S_\LetterConcSys$
     to  a global state $\gst' \in S_\LetterConcSys$ belongs to $R_\LetterConcSys$ iff there is
     an index $i$, $1\le i \le \syssize(\param)$, such that:
    \begin{description}
      \item[\trmove.]
          The $i$-th process \emph{moves}:
$(\gst[i], \gst'[i]) \in \restrict{R}{\param}$.
      \item[\trmaintain.]
    The values of the local variables of the other processes are preserved:
        for  every
        process index $j \ne i$, $1 \le j \le \syssize(\param)$, it
        holds that $\gst[j]
        \gleich{\{ \pc \} \cup \locset}  \gst'[j]$.
    \end{description}

    \item $\labelfun_\LetterConcSys: S_\LetterConcSys \rightarrow 2^{\Prop}$ is a state labeling function.
\end{itemize}

The set of global states $S_\LetterConcSys$ and the transition
     relation $R_\LetterConcSys$ are preserved under every
     transposition $i \leftrightarrow j$ of process indices $i$ and
     $j$ in $\{1, \dots, \syssize(\param)\}$.
That is, every system $\sConcSys$ is \emph{fully symmetric} by
     construction.

\paragraph{Temporal Logic.}

We specify properties of distributed algorithms in formulas of
     temporal logic \LTLX.
     An formula of \LTLX{} is defined inductively as:
\begin{itemize}
\item a literal $\aprop$ or $\neg\aprop$, where $\aprop \in \Prop$, or
\item $\ltlF \varphi$, $\ltlG \varphi$, $\varphi
     \ltlU \psi$, $\varphi \vee \psi$, and $\varphi \wedge \psi$,
     where $\varphi$ and $\psi$ are \LTLX\ formulas.
\end{itemize}

We use the standard definitions of paths and the semantics of the
     \LTLX\ formulas~\cite{CGP1999}.

\paragraph{Model checking an instance of a parameterized system.}

Now, we arrive at the formulation of a parameterized model checking
     problem.
Given:
\begin{itemize}
\item a domain $D$,
\item a parameterized process skeleton $\Sk = (S, S_0, R)$,
\item a resilience condition $\ResCond$ on parameters $\paraset$
  (generating a set of admissible parameters $\AdmP$),
\item parameter values $\param \in \AdmP$,
\item and an \LTLX\ formula $\varphi$,
\end{itemize}
check whether $\sConcSys  \models \varphi$.

\section{A Modeling Framework for Distributed Algorithms}\label{sec:DA}

In this section, we adapt the general definitions of the previous
     section to fault-tolerant distributed algorithms.
First we introduce atomic propositions that allow us to express
     typical specifications of distributed algorithms.
Then, we define our control flow automata (CFA) that are suitable to
     express threshold guarded distributed algorithms as parameterized
     process skeleton.

\subsection{Quantified Propositions for Distributed Algorithms}
\label{sec:APs}

We write specifications for our parameterized systems in \LTLX.
This contrasts the vast majority of work on parameterized model
     checking where \emph{indexed} temporal logics are used
     \cite{BCG1989,CTV2008,CTTV04,EN95}.
The reason for the use of indexed temporal logics is that they allow
     to express \emph{individual} process progress, e.g., in dining
     philosophers it is required that if a philosopher $i$ is hungry,
     then $i$ eventually eats.
Intuitively, dining philosophers
     requires us to trace indexed processes along a computation, e.g.,
     $\forall i.\; \ltlG  (\mbox{hungry}_i \rightarrow (\ltlF
     \mbox{eating}_i))$.

In  contrast, fault-tolerant distributed algorithms are typically used
     to achieve certain \emph{global} properties, as consensus
     (agreeing on a common value), or broadcast (ensuring that all
     processes deliver the same set of messages).
To capture these kinds of properties, we have to trace only
     existentially or universally quantified properties, e.g., part of
     the broadcast specification ({relay})~\cite{ST87:abc} states that
     if some correct process accepts a message, then all (correct)
     processes accept the message, that is, $(\ltlG ( \exists i.\;
     \mbox{accept}_i ))  \rightarrow (\ltlF (\forall j.\;
     \mbox{accept}_j))$.

We are therefore considering a temporal logic where the {\em
     quantification over processes is restricted to propositional
     formulas.} We will need two kinds of quantified propositional
     formulas.
First, we introduce the set $\PropPC$ that contains propositions that
     capture comparison against some status value $\pcval \in \PC$,
     i.e.,
$$
\left[\forall i.\; \pc_i = \pcval
     \right] \text{ and } \left[\exists i.\; \pc_i = \pcval \right].$$

This allows us to express specifications of distributed algorithms.
To express the mentioned relay property, we identify the status
     values where a process has accepted the message.
We may quantify over all processes as we will only model those
     processes explicitely that are restricted in their internal
     behavior, that is, correct or benign faulty processes.
More severe faults (e.g., Byzantine faults) are modeled via
     non-determinism.
For a detailed discussion see Section~\ref{sec:casestudy}.

Second, in order to express comparison of variables ranging over~$D$, we add a
     set of atomic propositions $\PropVAR$ that capture comparison of
     variables  $x$, $y$, and constant $c$ that all range over~$D$;
     $\PropVAR$ consists of  propositions of the form
$$\left[\exists i.\; x_i + c < y_i \right].$$

We then define $\Prop$ to be the disjoint union of $\PropPC$ and
     $\PropVAR$.
The labeling function $\labelfun_{\LetterConcSys}$ of a  system
     instance $\sConcSys$ maps its state~$\gst$ to expressions
     $\aprop$ from~$\Prop$ as follows:
\begin{align}
&\left[\forall i.\; \pc_i = \pcval \right] \in \labelfun_{\LetterConcSys}({\gst})
\text{ iff }
\Wedge{1 \le i \le \syssize(\param)}{}
\left(\gst[i].\pc = \pcval \right)
\nonumber\\
&\left[\exists i.\; \pc_i = \pcval \right] \in \labelfun_{\LetterConcSys}({\gst})
 \text{ iff }
\Vee{1 \le i \le \syssize(\param)}{}
\left(\gst[i].\pc = \pcval  \right)
\nonumber\\
&\left[\exists i.\; x_i + c < y_i \right] \in \labelfun_{\LetterConcSys}({\gst})
 \text{ iff }
\Vee{1 \le i \le \syssize(\param)}{}
\left(\gst[i].x + c < \gst[i].y \right)
\nonumber
\end{align}

\subsection{CFA for Threshold Guarded Distributed Algorithms}\label{sec:THBDA}

Processes that run distributed algorithms execute the same acyclic
     piece of code repeatedly.
In the parlance of distributed algorithms, a single execution of this
     code is called a step, and steps of correct processes are
     considered to be atomic.
Depending on the actual code, one can classify distributed algorithms
     by what may happen during a step.
For instance, in our case study, a step consists of a receive, a
     computation, and a sending phase.
Therefore, we are led to describe steps using the concept of  control
     flow automata (CFA), where paths from the initial to the final
     location of the CFA describe \emph{one step} of the distributed
     algorithm.

A \emph{control flow automata CFA} is a link-labeled directed acyclic
     graph $\CFA = (Q, q_I, q_F, E)$ with a finite set $Q$ of nodes,
     called the locations, an initial location $q_I\in Q$, and a final
     location $q_F\in Q$.
A path from $q_I$ to $q_F$ is used to describe one step of the
     distributed algorithm.
The edges have the form $E \subseteq Q \times \CFAOp \times Q$, where
     $\CFAOp$ is the set of operations whose \emph{syntax} is defined~as:

\begin{align}
\CFAvar ::=\; & \left< \text{name of a variable from } \Lambda \cup
\Gamma \right>\\
\CFAloc ::=\; & \left< \text{an element of }  \PC \right> \\
\CFAparam ::=\; &  \left< \text{name of a parameter variable from }
 \paraset \right>\\
\CFAcvar ::=\; & \CFAvar \mid \CFAdummy\\
\CFAparamcombi ::=\; & \CFAparam \mid int \mid \CFAparamcombi +
\CFAparamcombi \mid  \CFAparamcombi - \CFAparamcombi \\
\CFAthreshold ::=\; & \CFAparamcombi\label{cfa:defthreshold}\\
\CFAguard ::=\; & \CFApc = \CFAloc \mid \CFAthreshold \le \CFAvar  \mid
 \CFAguard \wedge \CFAguard \mid \neg \CFAguard \label{cfa:threshold}\\
\CFAcondatom ::=\; & \CFAcvar \le \CFAcvar + \CFAparamcombi \label{cfa:condatom}\\
\CFAcond ::=\; & \CFAcondatom \mid \CFAcond \wedge \CFAcond\\
\CFAOp ::=\; &  \CFApc := \CFAloc \mid \CFAinc \; \CFAvar \mid
\CFAguard \mid \CFAvar :=  \CFApickOp{\CFAdummy}{\CFApick}{\CFAcond}
\end{align}

In addition to constructs of standard control flow automata, we use the
     statement ``$\CFApickOp{\CFAdummy}{\CFApick}{\CFAcond}$''  that
     non-deterministically chooses a value $\CFAdummy$ that satisfies
     condition ``$\CFAcond$,'' if such a value exists, otherwise the
     statement blocks.
Moreover, there is a special variable $\pc$ ranging over $\PC$.
Most importantly, our threshold guarded commands can be expressed as
     combinations of threshold conditions via $\CFAguard$.

\paragraph{Operational semantics.}

To distinguish the notions of states in a process skeleton and states
     in a CFA, we call states in a CFA \emph{valuations} while states
     in process skeletons are called states.
The set of valuations are defined identically to the set of states
     defined in Section~\ref{sec:prelim} as $\PC \times D^{|\locset|}
     \times D^{|\globset|} \times D^{\numparam}$.
Then, the following shows the semantics where we denote by $v \models
     \CFAcond[v'.x / \CFAdummy]$ that $v$ models $\CFAcond$ if all
     occurrences of $\CFAdummy$ in $\CFAcond$ are replaced by $v'.x$:
\begin{align}
(v,v') \in \sem{\CFAguard}\;  \text{ iff } \;
&v \models \CFAguard  \;\wedge \; v'=v \label{sem:CFAguard}\\
(v,v') \in \sem{\CFApc := \CFAloc}\;  \text{ iff } \;
&v'.\CFApc = \CFAloc \; \wedge \;
v \gleich{\vars \setminus \{\pc\}} v'
  \label{sem:CFApcassign}\\
(v,v') \in \sem{\CFAinc \; x}\;  \text{ iff } \;
&v'.x = v.x + 1  \;\wedge \;
v \gleich{\vars \setminus \{x\}} v'
\label{sem:CFAinc} \\
(v,v') \in  \sem{x := \CFApickOp{\CFAdummy}{\CFApick}{\CFAcond}}
  \;  \text{ iff } \;
&v \models  \CFAcond[v'.x / \CFAdummy]
\; \wedge \;
v \gleich{\vars \setminus \{x\}} v'
\label{sem:CFApick}
\end{align}

In Section~\ref{sec:casestudy} we discuss how one can obtain a CFA
     from a description of a distributed algorithm based on pseudo
     code used in the literature.
Figure~\ref{Fig:STCFA} (page \pageref{Fig:STCFA}) provides the CFA
     that corresponds to the Algorithm~\ref{alg:ST87} we analyze in
     our case study.

\paragraph{Obtaining a process skeleton and a system instance from a CFA.}

Let us assume that $\PC$, $\PC_0$, $\locset$, $\globset$, $\paraset$,
     $\ResCond$, and $\syssize$ are given.
Given a CFA $\CFA$, we now define the process skeleton $\SkAut = (S,
     S^0, R)$ induced by~$\CFA$.

From the used variables and parameters we directly obtain
     the set $S$ of states.
We assume that all variables that range over $D$ are initialized
     to~$0$.
From this and $\PC_0$, we obtain~$S^0$.

It remains to define how the transition relation~$R$ is obtained from
     the semantics.
For two relations $R_1$ and $R_2$, we use the notation that $R_1\circ
     R_2 = \{ (x,z) \mid (x,y)\in R_1 \wedge (y,z)\in R_2\}$.
Each path in the CFA $\CFA$ from $q_I$ to $q_F$ induces a sequence of
     operations $\omega = o_1, \dots , o_k$ for some $k$; recall that
     the steps of a distributed algorithm are described by an acyclic
     CFA.
Then $\sem{\omega}$ is defined as $\sem{o_1} \circ \cdots \circ
     \sem{o_k}$, and the transition relation is defined by
$R = \bigcup_{\omega \text{ path in } \CFA}
     \sem{\omega}$.

We have thus defined the process skeleton $\SkAut$ induced by
     CFA~$\CFA$.
For a given $\param\in\AdmP$, a system instance  $\ConcSys$  is then
     the parallel composition of $\syssize(\param)$ process skeletons
     $\SkAut$, as defined in Section~\ref{sec:prelim}.

\medskip


\section{Transferring Pseudo-code to our Framework}\label{sec:casestudy}

\begin{algorithm}[t]
\footnotesize
\setlinenosize{\tiny}
\setlinenofont{\tt}
\begin{algorithmic}[1]
\CODE{$i$ if it is correct}
\VAR{}

\STATE $v_i\in \{ \false, \true \}$
\STATE $\accept_i\in\{ \false, \true  \}\leftarrow \false$
\EMPTY

\TRANS{}

\IF{$v_i$ {\bf and} not sent $\echomsg$ before}
\label{alg:init1}
\STATE \emph{send} $\echomsg$ to all;
\ENDIF

\IF{\emph{received} $\echomsg$ from at least $t+1$ \emph{distinct}
  processes\\ \qquad {\bf and} not sent $\echomsg$ before}
\label{alg:tp1}
\STATE \emph{send} $\echomsg$ to all;
\ENDIF

\IF{\emph{received} $\echomsg$ from at least $n-t$ \emph{distinct}
processes}
\label{alg:nmt}
\STATE $\accept_i \leftarrow \true$;\label{alg:end}
\ENDIF

\end{algorithmic}
\caption{ Core logic of the broadcasting algorithm from~\cite{ST87:abc}.}
\label{alg:ST87}
\end{algorithm}

We analyze Algorithm~\ref{alg:ST87}, which is the
     core of the broadcasting primitive by Srikanth and
     Toueg~\cite{ST87}.
In this section we first describe the computational model and
     Algorithm~\ref{alg:ST87} from a \emph{distributed algorithms
     point of view}, and will then show how to capture the algorithm
     in our modeling framework.

\paragraph{Computational model for asynchronous distributed
  algorithms.}

We recall the standard assumptions for asynchronous distributed
     algorithms.
As mentioned in the introduction, a system consists of $n$ processes
     out of which at most $t$ may be faulty.
When considering a fixed computation, we denote by $f$ the actual
     number of faulty processes.
It is assumed that $n>3t \wedge f \leq t \wedge t>0$.
 Correct processes
     follow the algorithm, in that they take steps that correspond to
     the algorithm description.
Between every pair of processes, there is a bidirectional link over
     which messages are exchanged.
A link contains two message buffers, each being the receive buffer of
     one of the incident processes.


A step of a correct process is atomic and consists of
     the following three parts.
First a process receives a possibly empty subset of the messages in
     its buffer, then it performs a state transition depending on its
     current state and the received messages.
Finally, a process may send at most one message to each process, that
     is, it puts a message in the buffer of the other processes.

Computations are  asynchronous in that the steps can be arbitrarily
     interleaved, provided that each correct process takes an infinite
     number of steps.
Moreover, if a message $m$ is put into a process $p$'s buffer, and $p$
     is correct, then~$m$ is eventually included in the set of
     messages received.
This property is called \emph{reliable communication}.
Faulty processes are not restricted, except that they have no influence
     of the buffers of links to which they are not incident.
This property is often called \emph{non-masquerading}, as a faulty
     process cannot ``pretend'' to be another process.

\input{CFAex.tex}

\paragraph{Specific details of Algorithm~\ref{alg:ST87}.}

The code is typical pseudocode found in the distributed algorithms
     literature.
The \codelinerange{alg:init1}{alg:end} describe one step of the
     algorithm.
Receiving messages is implicit and performed before
     \codelineref{alg:init1}, and the possible sending of messages is
     deferred to the end, and is performed after
     \codelineref{alg:end}.

We observe that a process always sends to all.
Moreover, \codelinerange{alg:init1}{alg:end} only
     consider messages of type $\echomsg$, while all other messages
     are ignored.
Hence, a Byzantine faulty process has an impact on
     correct processes only if they send an $\echomsg$ when they
     should not, or vice versa.
Note that faulty processes may behave two-faced, that is, send messages
     only to a subset of the correct processes.
Moreover, faulty processes may send multiple $\echomsg$ messages to a
     correct process.
However, from the code we observe that multiple receptions of such
     messages do not influence the number of messages received by
     ``distinct'' processes due to non-masquerading.
Finally, the condition ``not sent $\echomsg$ before'' guarantees that
     each correct process sends $\echomsg$ at most once.

\paragraph{Our modeling choices.}

The most immediate choice is that we consider the set of parameters
     $\paraset$ to be $\{n, t, f\}$ and $RC(n, t, f) = n>3t \wedge f
     \leq t \wedge t>0$.
In the pseudo code, the status of a process is only implicitly
     mentioned.
The relevant information we have to represent in the status variable
     is (i) the initial state (ii) whether a process has already sent
     $\echomsg$ and (iii) whether a process has set $\accept$ to
     $\true$.
Observe that once a process has sent $\echomsg$, its value of $v_i$
     does not interfere anymore with the further state transitions.
Moreover, a process only sets $\accept$ to $\true$ if it has sent a
     message (or is about to do so in the current step).
Hence, we define the set $\PC$ to be $\{\IT,\RI,\SE,\AC\}$, where
     $\PC_0= \{\IT,\RI \}$.
$\IT$ corresponds to the case where initially $v_i=\false$, and $\RI$
     to the case where initially $v_i=\true$.
Further, $\SE$ means that a process has sent an $\echomsg$ message but
     has not set $\accept$ to $\true$ yet, and $\AC$ means that the
     process has set $\accept$ to $\true$.
Having fixed the status values, we can formalize the specifications we
     want to verify.
They are obtained by the broadcasting specification parts called
     \emph{unforgeability}, \emph{correctness}, and \emph{relay}
     introduced in~\cite{ST87:abc}:
\begin{align}
  \tag{U}
 & \ltlG \left(\left[ \forall i.\; \pc_i \ne \RI \right]
     \rightarrow \ltlG
   \left[\forall j.\;  \pc_j \ne \AC
     \right]\right) \label{ST:u}
\\
  \tag{C}
&\ltlG \left(\left[\forall i.\; \pc_i = \RI \right]
      \rightarrow
   \ltlF  \left[ \exists j.\; \pc_j = AC\right]
   \right) \label{ST:corr}
 \\
\tag{R}
  &\ltlG \left(\left[\exists i.\; \pc_i = \AC \right]
      \rightarrow \ltlF \left[
   \forall j.\; \pc_j = AC \right] \right)
\label{ST:rel}
\end{align}
Note carefully that (\ref{ST:u}) is a safety specification while
     (\ref{ST:corr}) and (\ref{ST:rel}) are liveness specifications.

\smallskip

As the asynchrony of steps is already handled by our parallel
     composition described in Section~\ref{sec:prelim}, it remains to
     describe the semantics of sending and receiving messages in our
     system model using control flow automata.

Let us first focus on messages from and to correct processes.
As we have observed that each correct process sends at most one
     message, and multiple messages from faulty processes have no
     influence, it would be sufficient to represent each buffer by a
     single variable that represents whether a message of a certain
     kind has been put into the buffer.
As we have only $\echomsg$ messages sent by correct processes, it is sufficient
	 to model one variable per buffer.
Moreover, if we only consider the buffers between correct processes,
     due to the ``send to all'' it is sufficient to capture all
     messages between correct processes in a single variable.
To model this, we introduce the shared variable $\sent$.

The reception of messages can then be modeled by a local variable
     $\rcvd$ whose update depends on the messages sent.
In particular, upon a receive, the variable $\rcvd$ can be increased
     to any value less than or equal to $\sent$.

It remains to model faults.
As our system model is symmetric by construction, all processes must
     be identical processes.
This allows at least the two possibilities to model faults:
\begin{itemize}

\item we capture whether a process is correct or faulty as a flag in
     the status, and require that in each run $f\le t$
     processes are faulty.
Then we would have to derive a CFA sub-automaton for faulty processes,
     and would need additional variables to capture sent messages by
     faulty processes.

\item we consider the system to consist of correct processes only,
     let $\syssize(n,t,f) = n-f$, and model only the influence of
     faults, via the messages correct processes may receive.
This can be done by allowing each correct process to receive at most
     $f$ messages more than sent by correct ones, that is that $\rcvd$
     can be increased to any value less than or equal to $\sent + f$.

\end{itemize}

Implementing the first option would require more variables, namely,
     the additional flag to distinguish correct from faulty processes,
     and the additional variables to capture messages by faulty
     processes.
These variables would increase the state space, and would make this
     option non-practical.
Moreover, we would have to capture the number of faults $f$, and the
     corresponding resilience condition.
Therefore, we have implemented the latter approach for our experiments
     in Section~\ref{sec:exp}.

Based on this discussion we directly obtain the CFA given in
     Figure~\ref{Fig:STCFA} that describes the steps of
     Algorithm~\ref{alg:ST87}.
Note that its structure follows the pseudo code description of
     Algorithm~\ref{alg:ST87} very closely.

\paragraph{Verification strategy for liveness.}
\label{sec:veriDA}

Relevant liveness properties can typically only be guaranteed if the
     underlying system ensures some fairness guarantees.
In asynchronous distributed systems one assumes for instance
     communication fairness, that is, every message sent is eventually
     received.
The statement $\exists i.\; \rcvd_i < \sent_i$ describes a
     global state where messages are still in transit.
It follows that a formula $\psi$ defined by
\begin{equation}
 \ltlF \ltlG \left[\exists i.\;
     \rcvd_i < \sent_i \right]  \tag{inequity} \label{F:RC}
\end{equation}
 states that the system violates
     communication fairness.
We only require a liveness specification $\varphi$ to be
     satisfied if the system is communication fair. In other words,
     $\varphi$ is satisfied \emph{or} the communication is unfair,
     that is, $\varphi \vee \psi$.
Our approach is to automatically verify $\varphi \vee \psi$.

\smallskip

Along all paths where communication is fair, the value of
     $\rcvd_i$ has at least to reach the value of $\sent_i$.
Since $\rcvd_i$ can only increase upon a step by $i$, $i$ is forced
     to take steps as long as it has not received messages yet.
That is, by this modeling, communication fairness implies some form of
     computation fairness.


\paragraph{Modeling other fault scenarios.}

Fault scenarios other than Byzantine faults can be modeled by changing
     the system size, using conditions similar to (\ref{F:RC}), and
     slightly changing the CFA.
More precisely, by changing the non-deterministic assignment (the edge
     leaving $q_I$) that corresponds to receiving messages.
For instance, replacing Byzantine by send omission process
     faults~\cite{NT90} could be modeled as follows: Faulty processes
     could be modeled explicitly by setting $\syssize(n,t,f) = n$.
That at most $f$ processes may fail to send messages, could be modeled
     by $ \ltlF \ltlG \left[\exists i.\; \rcvd_i + f < \sent_i
     \right]$.
Finally, in this fault model processes may receive all messages sent,
     that is,  $\rcvd :=
     \CFApickOp{\CFAdummy}{\CFApick}{\rcvd\le\CFAdummy  \; \wedge \;
     \CFAdummy \le\sent}$.
By similar adaptations one models, e.g., corrupted communication
     (e.g., due to value faulty links) \cite{SW89}, or hybrid fault
     models \cite{BSW11:hyb} that contain different fault scenarios.

\section{Experimental Evaluation}\label{sec:exp}

\begin{figure}[p]
\begin{center}
\begin{minipage}{.8\linewidth}
  \begin{lstlisting}[language=promela,label=Lst:woswasi]
symbolic int N, T, F;
assume(N > 3 && F >= 0 && T >= 1);
assume(N > 3 * T && F <= T);

atomic ex_acc = some(Proc:pc == AC);
atomic all_acc = all(Proc:pc == AC);

atomic in_transit = some(Proc:nrcvd < nsnt);

active[N - F] proctype Proc() {
    byte pc = 0, next_pc = 0;
    int nrcvd = 0, next_nrcvd = 0;
[...]
  do
    :: atomic {
[...]
      if
        :: next_nrcvd >= N - T ->
          next_pc = AC;
        :: next_nrcvd < N - T &&
           (pc == V1 || next_nrcvd >= T + 1) ->
          next_pc = SE;
        :: else ->
          next_pc = pc;
      fi;
      /* send the echo message */
      if
        :: (pc == V0 || pc == V1) &&
            (next_pc == SE || next_pc == AC) ->
          nsnt++;
        :: else;
      fi;
[...]
     }
  od;
}
  \end{lstlisting}
\end{minipage}
\end{center}
\caption{Example encoding of the CFA in our Promela extension.}
\label{Fig:promela}
\end{figure}

We have extended Spin's~\cite{H2003} input language Promela to be able
     to express our control flow automata that operate on unbounded
     variables and symbolic variables to express parameters.
Figure~\ref{Fig:promela} provides the central parts of the code of our
     case study.
For the experiments we have implemented four distributed  algorithms
     that use threshold guarded commands.
They differ in the guarded commands, and work for different fault
     assumptions.
The following list is ordered from the most general fault model to the
     most restricted one.
The given resilience conditions on $n$ and $t$ are the ones we
     expected from the literature, and their tightness was confirmed
     by our experiments:
\begin{description}
\item[\textsc{\sc Byz}.] tolerates $t$ Byzantine faults if $n>3t$,
\item[\textsc{\sc symm}.]  tolerates $t$ symmetric (identical
     Byzantine~\cite{AW04}) faults if $n>2t$,
\item[\textsc{\sc omit}.] tolerates $t$ send omission
     faults if $n>2t$,
\item[\textsc{\sc clean}.] tolerates $t$ clean crash faults for $n>t$.
\end{description}
The CFAs of these algorithms follow the same principles, so we do not give all
     of them in this paper.
Figure~\ref{Fig:STCFA} provides the most complicated one, namely
     \textsc{Byz} (we discussed how it is obtained from the literature
     in detail in Section~\ref{sec:casestudy}), next to the CFA of
     \textsc{clean} which actually is the simplest one.
Our tool takes as input a CFA encoded in extended Promela, and
     concrete values for parameters, generates as output standard
     Promela code.

\begin{table}[t]
 \begin{center}
  \begin{tabular}{|ll|cc|r|r|r|r|r|}
   \hline
\textbf{\#}  & \textbf{\scriptsize{parameter values}} &
\textbf{\scriptsize{spec}} & \textbf{\scriptsize{valid}} &
\textbf{\scriptsize{Time}} & \textbf{\scriptsize{Mem.}} &
\textbf{\scriptsize{Stored}} & \textbf{\scriptsize{Transitions}} &
\textbf{\scriptsize{Depth}} \\
   \hline
\multicolumn{9}{c}{{\sc Byz}}\\
 \hline
\textbf{B1}  & N=7,T=2,F=2 & (\ref{ST:u}) & \checkmark & 3.13 sec. & 74 MB & $193 \cdot 10^3$ & $1 \cdot 10^6$ & 229 \\
   \hline
\textbf{B2}  & N=7,T=2,F=2 & (\ref{ST:corr}) & \checkmark & 3.43 sec. & 75 MB & $207 \cdot 10^3$ & $2 \cdot 10^6$ & 229 \\
   \hline
\textbf{B3}  & N=7,T=2,F=2 & (\ref{ST:rel}) & \checkmark & 6.3 sec. & 77 MB & $290 \cdot 10^3$ & $3 \cdot 10^6$ & 229 \\
   \hline
\textbf{B4}  & N=7,T=3,F=2 & (\ref{ST:u}) & \checkmark & 4.38 sec. & 77 MB & $265 \cdot 10^3$ & $2 \cdot 10^6$ & 233 \\
   \hline
\textbf{B5}  & N=7,T=3,F=2 & (\ref{ST:corr}) & \checkmark & 4.5 sec. & 77 MB & $271 \cdot 10^3$ & $2 \cdot 10^6$ & 233 \\
   \hline
\textbf{B6}  & N=7,T=3,F=2 & (\ref{ST:rel}) & \xmark & 0.02 sec. & 68 MB & $1 \cdot 10^3$ & $13 \cdot 10^3$ & 210 \\
   \hline
\multicolumn{9}{c}{{\sc omit}}\\
 \hline
\textbf{O1}  & N=5,To=2,Fo=2 & (\ref{ST:u}) & \checkmark & 1.43 sec. & 69 MB & $51 \cdot 10^3$ & $878 \cdot 10^3$ & 175 \\
   \hline
\textbf{O2}  & N=5,To=2,Fo=2 & (\ref{ST:corr}) & \checkmark & 1.64 sec. & 69 MB & $60 \cdot 10^3$ & $1 \cdot 10^6$ & 183 \\
   \hline
\textbf{O3}  & N=5,To=2,Fo=2 & (\ref{ST:rel}) & \checkmark & 3.69 sec. & 71 MB & $92 \cdot 10^3$ & $2 \cdot 10^6$ & 183 \\
   \hline
\textbf{O4}  & N=5,To=2,Fo=3 & (\ref{ST:u}) & \checkmark & 1.39 sec. & 69 MB & $51 \cdot 10^3$ & $878 \cdot 10^3$ & 175 \\
   \hline
\textbf{O5}  & N=5,To=2,Fo=3 & (\ref{ST:corr}) & \xmark & 1.63 sec. & 69 MB & $53 \cdot 10^3$ & $1 \cdot 10^6$ & 183 \\
   \hline
\textbf{O6}  & N=5,To=2,Fo=3 & (\ref{ST:rel}) & \xmark & 0.01 sec. & 68 MB & 17 & 135 & 53 \\
   \hline
\multicolumn{9}{c}{{\sc symm}}\\
 \hline
\textbf{S1}  & N=5,T=1,Fp=1,Fs=0 & (\ref{ST:u}) & \checkmark & 0.04 sec. & 68 MB & $3 \cdot 10^3$ & $23 \cdot 10^3$ & 121 \\
   \hline
\textbf{S2}  & N=5,T=1,Fp=1,Fs=0 & (\ref{ST:corr}) & \checkmark & 0.03 sec. & 68 MB & $3 \cdot 10^3$ & $24 \cdot 10^3$ & 121 \\
   \hline
\textbf{S3}  & N=5,T=1,Fp=1,Fs=0 & (\ref{ST:rel}) & \checkmark & 0.08 sec. &
68 MB & $5 \cdot 10^3$ & $53 \cdot 10^3$ & 121 \\
   \hline
\textbf{S4}  & N=5,T=3,Fp=3,Fs=1 & (\ref{ST:u}) & \checkmark & 0.01 sec. & 68 MB & 66 & 267 & 62 \\
   \hline
\textbf{S5}  & N=5,T=3,Fp=3,Fs=1 & (\ref{ST:corr}) & \xmark & 0.01 sec. & 68 MB & 62 & 221 & 66 \\
   \hline
\textbf{S6}  & N=5,T=3,Fp=3,Fs=1 & (\ref{ST:rel}) & \checkmark & 0.01 sec. & 68 MB & 62 & 235 & 62 \\
   \hline
\multicolumn{9}{c}{{\sc clean}}\\
 \hline
\textbf{C1} & N=3,Tc=2,Fc=2,Fnc=0 & (\ref{ST:u}) & \checkmark & 0.01 sec. & 68 MB & 668 & $7 \cdot 10^3$ & 77 \\
   \hline
\textbf{C2} & N=3,Tc=2,Fc=2,Fnc=0 & (\ref{ST:corr}) & \checkmark & 0.01 sec. & 68 MB & 892 & $8 \cdot 10^3$ & 81 \\
   \hline
\textbf{C3} & N=3,Tc=2,Fc=2,Fnc=0 & (\ref{ST:rel}) & \checkmark & 0.02 sec. & 68 MB & $1 \cdot 10^3$ & $17 \cdot 10^3$ & 81 \\
   \hline
  \end{tabular}
 \end{center}
\caption{Summary of experiments}\label{tab:exp}
\end{table}

\begin{figure*}[t]
    \begin{minipage}{.5\linewidth}
        \includegraphics[width=1\linewidth]{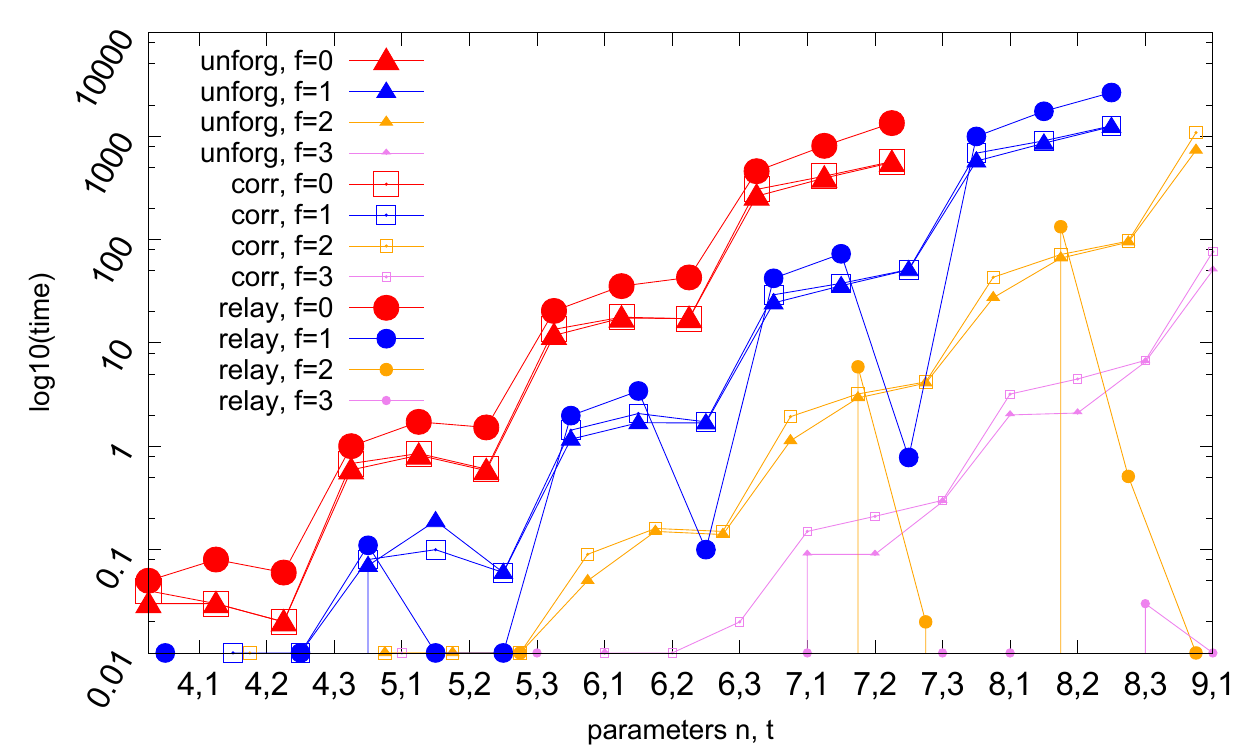}
        \caption{Spin running time on \textsc{byz}.}
        \label{fig:byz-time}
    \end{minipage}
    \begin{minipage}{.5\linewidth}
        \includegraphics[width=1\linewidth]{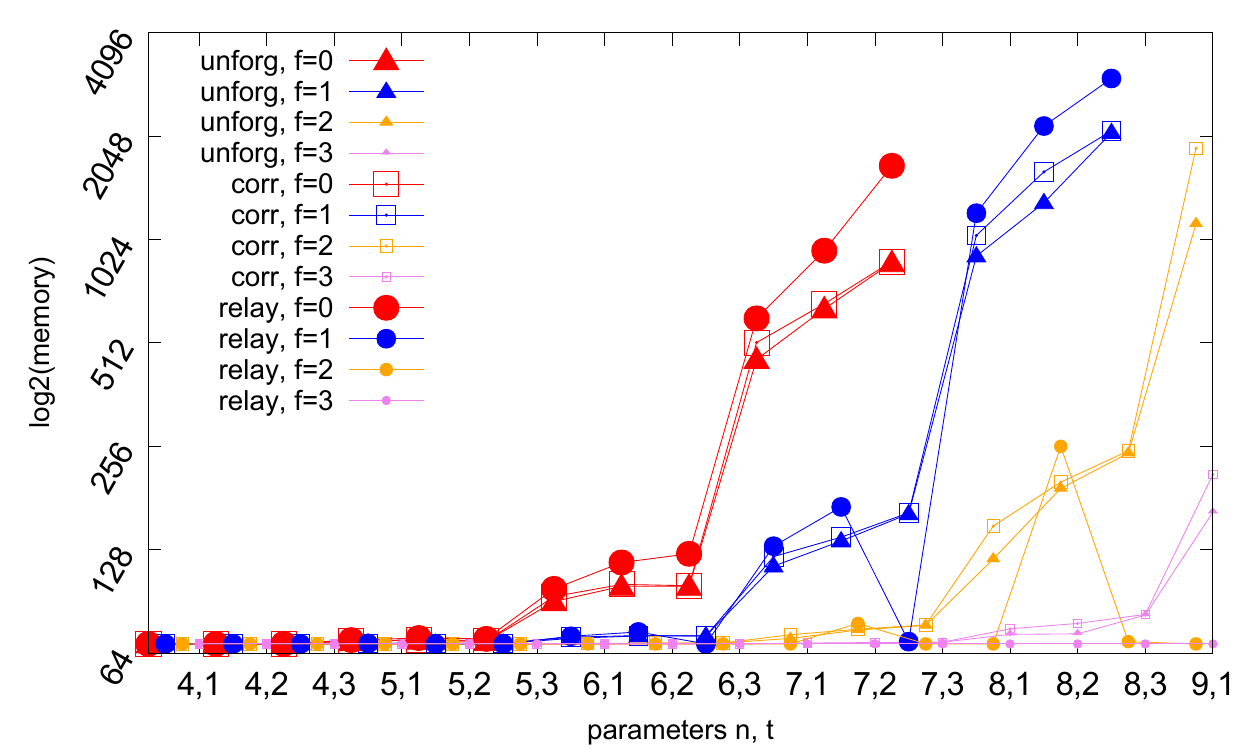}
        \caption{Spin memory usage on \textsc{byz}.}
        \label{fig:byz-mem}
    \end{minipage}
\end{figure*}

The major goal of the experiments was to check the adequacy of our
     formalization.
To this end we considered the four mentioned well-understood
     distributed algorithms.
For each of which we systematically\footnote{Complete experimental
     data is given in the appendix.} changed the parameter values, in
     order to ascertain that under our modeling, the different
     combination of parameters lead to the expected result.
     Table~\ref{tab:exp} and Figures~\ref{fig:byz-time},~\ref{fig:byz-mem}
     summarize the results of our experiments.

Lines~B1\ndash---B3, O1\ndash---O3, S1\ndash---S3, and C1\ndash---C3
     capture the cases that are within the resilience condition known
     for the respective algorithm, and the algorithms were verified by
     Spin.
In Lines~B4\ndash---B6, the algorithm's parameters are chosen to
     achieve a goal that is known to be impossible~\cite{LSP80}, i.e.,
     to tolerate that 3 out of 7 processes may fail.
This violates the $n>3t$ requirement.
Our experiment shows that even if only 2 faults occur in this setting,
     the relay specification (\ref{ST:rel}) is violated.
In Lines~O4\ndash---O6, the algorithm is designed properly, i.e., 2
     out of 5 processes may fail ($n>2t$ in the case of omission
     faults).
Our experiments show that this algorithm fails in the presence of 3
     faulty processes, i.e., (\ref{ST:corr}) and (\ref{ST:rel}) are
     violated.

For slightly bigger systems, that is, for $n=11$ our experiments run
     out of memory.
This shows the need for parameterized verification of these
     algorithms.

\section{Related Work}\label{sec:relwork}

In the area of verification, the most closest work to ours
     is~\cite{FismanKL08} which introduces a framework that targets at
     parameterized fault-tolerant distributed algorithms.
However, \cite{FismanKL08} only considers fixed size (and finite)
     process descriptions which, consequently, cannot depend on the
     parameters $n$, $t$, and $f$.
This makes it impossible for the algorithms to use thresholds for two
     reasons: (i) With fixed size variables it is impossible to count
     messages in a parameterized setting.
(ii) When process descriptions do not refer to parameters, it is
     impossible to compare counters against e.g., the parameter $t$, a
     standard construction in distributed algorithms.
While \cite{FismanKL08} contains ideas to model faults, the formalism
     does only allow to express very limited fault-tolerant
     distributed algorithms.
For instance, their verification example considered a broadcasting
     algorithm in the case of crash faults, that has a trivial
     threshold guard, namely where one checks whether one message is
     received.
As explained in the introduction, these kinds of rules are problematic
     in the presence of more severe fault types as Byzantine faults.
Finally, the experimental data they provide is restricted to reporting
     that for $f=17$ their algorithm was verified.
The algorithm they considered as a case study works in a very simple
     setting, namely it is correct for all combinations of $n$ and
     $t$, so they did not have to consider resilience conditions.
However, failure models that are more involved than their crash
     assumptions typically call for special constraints on $n$ and
     $t$.
Moreover, they use a specification of reliable broadcast which differs
     from the distributed algorithms literature (e.g., \cite{HT93}).
In fact, their specification can be satisfied with a trivial algorithm
     (consisting of a single assignment $\Sigma_v := \bot$).

To conclude, \cite{FismanKL08} considered the very important area of
     formal verification of fault-tolerant distributed algorithms, and
     showed what kind of modeling is feasible with techniques from
     regular model checking.
From the points discussed above, it is clear the their approach
     although making interesting progress falls short of several
     aspects of fault-tolerant distributed algorithms.
An important goal of our work is to initiate a systematic study of
     distributed algorithms from a verification and programming
     language point of view in a way that does not betray the
     fundamentals of distributed algorithms.
We believe that ours is indeed the first paper about model checking of
     an adequately modeled fault-tolerant distributed algorithm.
In our companion paper~\cite{JKSVW12a}, we even show how to verify the
     algorithms from the experiments above for all system sizes, and
     we thus actually verify the algorithms rather than just instances
     of the algorithms.

\smallskip

The I/O Automata framework \cite{LT89,2006Kaynar,MitraL07} models a
     distributed system as a collection of automata representing
     processes and of automata representing the communication medium,
     e.g., message passing links.
The framework concentrates on the interfaces\dash---input and output
     actions\dash---rather than on semantics, and in fact much of the
     IOA literature on distributed algorithms uses pseudo code to
     describe what happens, e.g., upon an input event.
In contrast to I/O Automata, which focus on the interfaces between
     processes (input, output), our CFAs focus on the semantics for
     steps of distributed algorithms, and the construction of a system
     instance as a Kripke structure corresponds directly to standard
     distributed computing models like \cite{FLP85,DDS87,DLS88}, that
     are build around \emph{steps} rather than input or output
     \emph{actions}.

\smallskip

The \emph{temporal logic of actions} \cite{Lamport2002} is a variant
     of temporal logic by \cite{Pnueli:1977}, and built upon it, TLA+
     is a specification language for concurrent and reactive systems.
This approach is very general, because one can express a wide variety
     of systems in TLA and TLA+.
As our domain-specific framework is built specifically for distributed
     algorithms, we focused on their specifics such as resilience
     conditions, faults, and asynchrony.

\section{Conclusions}

We introduced a framework to capture threshold-based fault-tolerant
     distributed algorithms.
The framework consists of a parametric system model and of control
     flow automata, which allows to express the non-determinism
     typical for distributed algorithms.
We explained in detail how an algorithm from the literature can be
     formalized in this framework.

We verified the appropriateness of our modeling by model checking four
     well-understood fault-tolerant distributed algorithms for fixed
     system sizes.
This shows that the framework is a starting point to address many
     exciting verification problems in the area of distributed
     algorithms.
In fact, in a companion paper~\cite{JKSVW12a} is dealing with the
     parameterized verification problem.
There, we are using the framework to apply several abstraction
     techniques which allowed us to verify the four algorithms for all
     combinations of parameters admitted by the resilience condition.

\bibliographystyle{splncs03}

\bibliography{lit,dialog}





\newpage

\appendix

\section*{APPENDIX}

\section{Experimental data}

This section provides a complete set of experiments.
The highlighted lines are the ones we have chosen for the body of the
     manuscript.

\input{t-symm-concrete}

\newpage

\input{t-clean-concrete}

\input{t-byz-concrete}

\input{t-omit-concrete}

\begin{figure*}[t]
    \begin{minipage}{.5\linewidth}
        \includegraphics[width=1\linewidth]{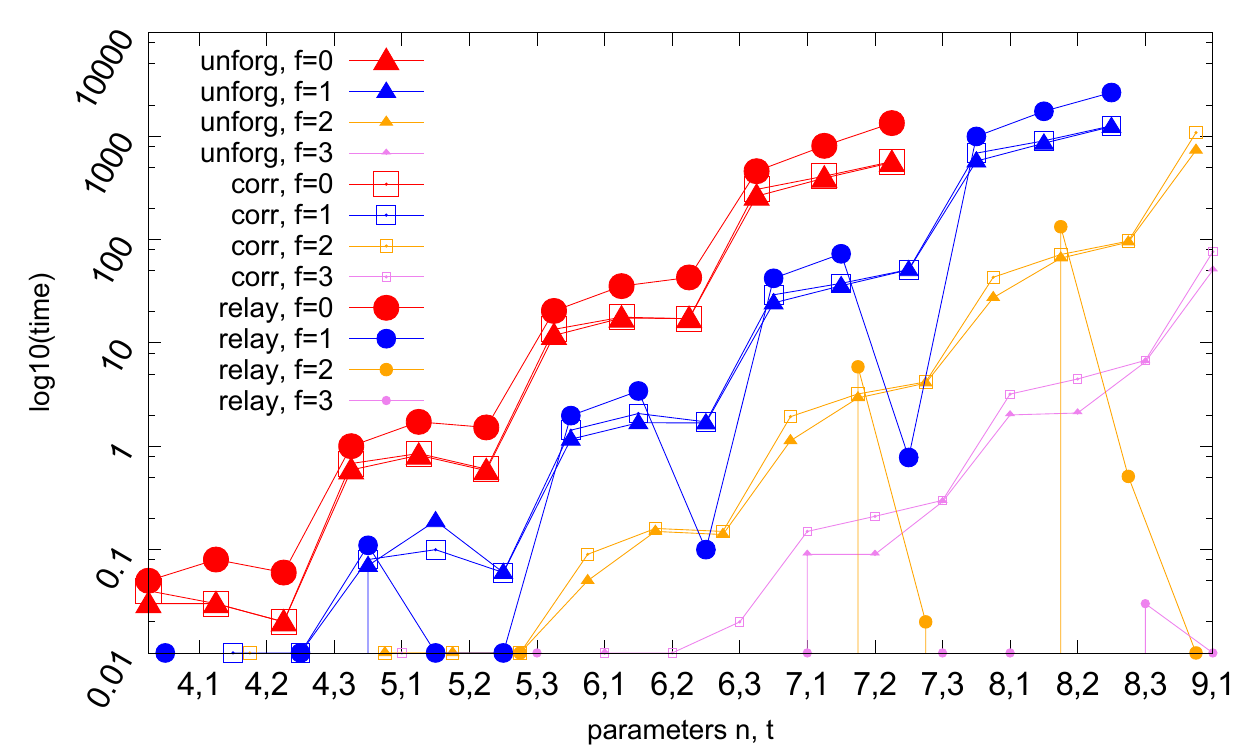}
        \caption{Spin running time on \textsc{byz}.}
    \end{minipage}
    \begin{minipage}{.5\linewidth}
        \includegraphics[width=1\linewidth]{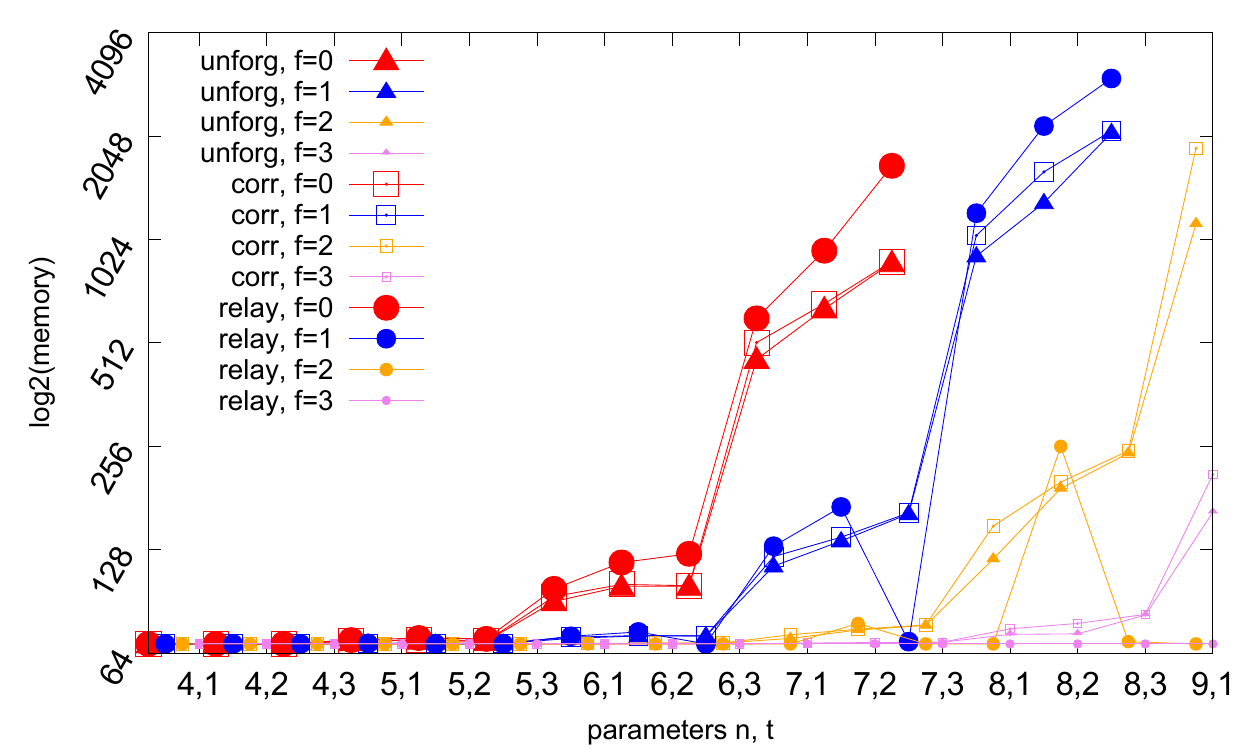}
        \caption{Spin memory usage on \textsc{byz}.}
    \end{minipage}
\end{figure*}

\end{document}

%% file: CFAex.tex
\tikzstyle{node}=[circle,draw=black,thick,minimum size=4mm,font=\small]
\tikzstyle{sink}=[circle,draw=black,fill=black!10,very thick,minimum size=4mm]
\tikzstyle{dest}=[circle,draw=black!50,fill=black!20,thick,minimum size=4mm,font=\small]
\tikzstyle{post}=[->,thick]
\tikzstyle{pre}=[<-,thick]
\tikzstyle{cond}=[rounded
  corners,rectangle,minimum
  width=1cm,fill=white,draw=black,font=\small]
\tikzstyle{asign}=[rectangle,minimum
  width=1cm,fill=black!5,draw=black,font=\small]

\begin{figure}[t]
\begin{minipage}{.5\linewidth}
\begin{center}
\scalebox{.9}{
\begin{tikzpicture}[>=latex,rotate=0]

  \node at ( 0,11.6) [dest] (I) {$q_I$};

  \node at ( 0,10) [node] (D) {$q_1$};
  \node at ( 3,9.5) [node] (1) {$q_2$} ;
  \node at ( 3,7.5) [node] (2) {$q_3$};
  \node at ( 0,7) [node] (3) {$q_4$};  
\draw [post] (D) -- node[cond] {$\CFApc = \RI$} (1);
\draw [post] (D) -- node[cond] {$\neg(\CFApc = \RI$)} (3);
\draw [post] (1) -- node[asign] {$\CFAinc \; \sent$} (2);
\draw [post] (2) -- node[asign] {$\CFApc := \SE$} (3);

  \node at ( 3,6) [node] (4) {$q_{5}$};
  \node at ( 1,4.5) [node] (5) {$q_6$};
  \node at ( 3,3) [node] (6) {$q_7$} ;

  \node at ( 1,1.5) [node] (7) {$q_8$};

  \node at ( 3,0.25) [node] (8) {$q_9$};
  \node at ( 0,0) [dest] (9) {$q_F$};

\draw [post] (I) --  (D);
\node at ( 1.5,10.9) [asign] {$\rcvd := \CFApickOp{\CFAdummy}{\CFApick}{\rcvd\le\CFAdummy  \; \wedge \;$ $\CFAdummy \le\sent+f}$};

\draw [post] (3) -- node[cond] {$\neg(t+1 \le \rcvd)$}  (9);  
\draw [post] (3) -- node[cond] {$t+1\le\rcvd$} (4);

\draw [post] (4) -- node[cond] {$\CFApc = \IT$} (5);
\draw [post] (4) -- node[cond] {$\neg(\CFApc = \IT)$} (6);
\draw [post] (5) -- node[asign] {$\CFAinc \; \sent$} (6);

\draw [post] (6) -- node[cond] {$n-t \le \rcvd$} (7); 
\draw [post] (6) -- node[cond] {$\neg(n-t \le \rcvd)$} (8); 

\draw [post] (8) -- node[asign] {$\CFApc := \SE$} (9);
\draw [post] (7) -- node[asign,right=-0.25cm] {$\CFApc := \AC$} (9);

\end{tikzpicture}
}
\end{center}
\end{minipage}
\begin{minipage}{.5\linewidth}
\begin{center}
\scalebox{0.9}{
\begin{tikzpicture}[>=latex,rotate=0]

  \node at ( 0,11.6) [dest] (I) {$q_I$};

  \node at ( 0,10) [node] (D) {$q_1$};
  \node at ( 3,9.5) [node] (1) {$q_2$} ;
  \node at ( 3,7.5) [node] (2) {$q_3$};
  \node at ( 0,7) [node] (3) {$q_4$};  
\draw [post] (D) -- node[cond] {$\CFApc = \RI$} (1);
\draw [post] (D) -- node[cond] {$\neg(\CFApc = \RI$)} (3);
\draw [post] (1) -- node[asign] {$\CFAinc \; \sent$} (2);
\draw [post] (2) -- node[asign] {$\CFApc := \SE$} (3);

  \node at ( 3,3.5) [node] (4) {$q_{5}$};

  \node at ( 0,0) [dest] (9) {$q_F$};

\draw [post] (I) --  (D);
\node at ( 1.5,10.9) [asign] {$\rcvd := \CFApickOp{\CFAdummy}{\CFApick}{\rcvd\le\CFAdummy  \; \wedge \;$ $\CFAdummy \le\sent}$};

\draw [post] (3) -- node[cond] {$\neg(n-t \le \rcvd)$}  (9);  
\draw [post] (3) -- node[cond] {$n-t\le\rcvd$} (4);

\draw [post] (4) -- node[cond] {$\CFApc := \AC$} (9);

\end{tikzpicture}
}
\end{center}
\end{minipage}
\caption{Control flow automaton for the steps of 
Algorithm~\ref{alg:ST87} on the left. On the right a CFA for a
distributed algorithm tolerating clean crash faults.}
\label{Fig:STCFA}
\end{figure}

%% file: t-symm-concrete.tex
  \begin{longtable}{llllllllllllllll}
   \hline
\textbf{\#} & \textbf{\scriptsize{param}} & \textbf{\scriptsize{spec}} & \textbf{\scriptsize{valid}} & \textbf{\scriptsize{SpinTime}} & \textbf{\scriptsize{SpinMemory}} & \textbf{\scriptsize{Stored}} & \textbf{\scriptsize{Transitions}} & \textbf{\scriptsize{Depth}} \\
   \hline
\textbf{1}  & N=3,T=1,Fs=1,Fp=0 & unforg & \checkmark & 0.01 sec. & 68.019 MB & 533 & $3 \cdot 10^3$ & 89 \\
   \hline
\textbf{2}  & N=3,T=1,Fs=1,Fp=0 & corr & \checkmark & 0.01 sec. & 68.019 MB & 578 & $3 \cdot 10^3$ & 89 \\
   \hline
\textbf{3}  & N=3,T=1,Fs=1,Fp=0 & relay & \checkmark & 0.01 sec. & 68.019 MB & 826 & $6 \cdot 10^3$ & 89 \\
   \hline
\textbf{4}  & N=5,T=1,Fp=0,Fs=0 & unforg & \checkmark & 0.6 sec. & 69.191 MB & $39 \cdot 10^3$ & $375 \cdot 10^3$ & 177 \\
   \hline
\textbf{5}  & N=5,T=1,Fp=0,Fs=0 & corr & \checkmark & 0.62 sec. & 69.191 MB & $39 \cdot 10^3$ & $379 \cdot 10^3$ & 177 \\
   \hline
\textbf{6}  & N=5,T=1,Fp=0,Fs=0 & relay & \checkmark & 1.56 sec. & 70.363 MB & $70 \cdot 10^3$ & $976 \cdot 10^3$ & 177 \\
   \hline
\rowcolor{gray}\textbf{7}  & N=5,T=1,Fp=1,Fs=0 & unforg & \checkmark & 0.04 sec. & 68.019 MB & $3 \cdot 10^3$ & $23 \cdot 10^3$ & 121 \\
   \hline
\rowcolor{gray}\textbf{8}  & N=5,T=1,Fp=1,Fs=0 & corr & \checkmark & 0.03 sec. & 68.019 MB & $3 \cdot 10^3$ & $24 \cdot 10^3$ & 121 \\
   \hline
\rowcolor{gray}\textbf{9}  & N=5,T=1,Fp=1,Fs=0 & relay & \checkmark & 0.08 sec. & 68.019 MB & $5 \cdot 10^3$ & $53 \cdot 10^3$ & 121 \\
   \hline
\textbf{10}  & N=5,T=1,Fp=1,Fs=1 & unforg & \checkmark & 0.06 sec. & 68.215 MB & $5 \cdot 10^3$ & $45 \cdot 10^3$ & 139 \\
   \hline
\textbf{11}  & N=5,T=1,Fp=1,Fs=1 & corr & \checkmark & 0.06 sec. & 68.215 MB & $6 \cdot 10^3$ & $45 \cdot 10^3$ & 139 \\
   \hline
\textbf{12}  & N=5,T=1,Fp=1,Fs=1 & relay & \checkmark & 0.16 sec. & 68.215 MB & $10 \cdot 10^3$ & $111 \cdot 10^3$ & 139 \\
   \hline
\textbf{13}  & N=5,T=2,Fp=0,Fs=0 & unforg & \checkmark & 1.04 sec. & 69.777 MB & $53 \cdot 10^3$ & $515 \cdot 10^3$ & 188 \\
   \hline
\textbf{14}  & N=5,T=2,Fp=0,Fs=0 & corr & \checkmark & 0.89 sec. & 69.777 MB & $55 \cdot 10^3$ & $533 \cdot 10^3$ & 188 \\
   \hline
\textbf{15}  & N=5,T=2,Fp=0,Fs=0 & relay & \checkmark & 1.81 sec. & 70.754 MB & $83 \cdot 10^3$ & $1 \cdot 10^6$ & 188 \\
   \hline
\textbf{16}  & N=5,T=2,Fp=1,Fs=0 & unforg & \checkmark & 0.04 sec. & 68.019 MB & $3 \cdot 10^3$ & $29 \cdot 10^3$ & 131 \\
   \hline
\textbf{17}  & N=5,T=2,Fp=1,Fs=0 & corr & \checkmark & 0.05 sec. & 68.019 MB & $4 \cdot 10^3$ & $33 \cdot 10^3$ & 131 \\
   \hline
\textbf{18}  & N=5,T=2,Fp=1,Fs=0 & relay & \checkmark & 0.08 sec. & 68.019 MB & $5 \cdot 10^3$ & $49 \cdot 10^3$ & 131 \\
   \hline
\textbf{19}  & N=5,T=2,Fp=1,Fs=1 & unforg & \checkmark & 0.09 sec. & 68.215 MB & $7 \cdot 10^3$ & $61 \cdot 10^3$ & 149 \\
   \hline
\textbf{20}  & N=5,T=2,Fp=1,Fs=1 & corr & \checkmark & 0.09 sec. & 68.215 MB & $8 \cdot 10^3$ & $62 \cdot 10^3$ & 149 \\
   \hline
\textbf{21}  & N=5,T=2,Fp=1,Fs=1 & relay & \checkmark & 0.18 sec. & 68.41 MB & $12 \cdot 10^3$ & $125 \cdot 10^3$ & 149 \\
   \hline
\textbf{22}  & N=5,T=2,Fp=2,Fs=0 & unforg & \checkmark & 0.01 sec. & 68.019 MB & 323 & $1 \cdot 10^3$ & 86 \\
   \hline
\textbf{23}  & N=5,T=2,Fp=2,Fs=0 & corr & \checkmark & 0.01 sec. & 68.019 MB & 412 & $2 \cdot 10^3$ & 86 \\
   \hline
\textbf{24}  & N=5,T=2,Fp=2,Fs=0 & relay & \checkmark & 0.01 sec. & 68.019 MB & 359 & $2 \cdot 10^3$ & 88 \\
   \hline
\textbf{25}  & N=5,T=2,Fp=2,Fs=1 & unforg & \checkmark & 0.01 sec. & 68.019 MB & 662 & $3 \cdot 10^3$ & 98 \\
   \hline
\textbf{26}  & N=5,T=2,Fp=2,Fs=1 & corr & \checkmark & 0.01 sec. & 68.019 MB & 762 & $4 \cdot 10^3$ & 98 \\
   \hline
\textbf{27}  & N=5,T=2,Fp=2,Fs=1 & relay & \checkmark & 0.01 sec. & 68.019 MB & 856 & $6 \cdot 10^3$ & 98 \\
   \hline
\textbf{28}  & N=5,T=2,Fp=2,Fs=2 & unforg & \checkmark & 0.01 sec. & 68.019 MB & $1 \cdot 10^3$ & $6 \cdot 10^3$ & 110 \\
   \hline
\textbf{29}  & N=5,T=2,Fp=2,Fs=2 & corr & \checkmark & 0.01 sec. & 68.019 MB & $1 \cdot 10^3$ & $6 \cdot 10^3$ & 110 \\
   \hline
\textbf{30}  & N=5,T=2,Fp=2,Fs=2 & relay & \checkmark & 0.02 sec. & 68.019 MB & $1 \cdot 10^3$ & $11 \cdot 10^3$ & 110 \\
   \hline
\textbf{31}  & N=5,T=3,Fp=0,Fs=0 & unforg & \checkmark & 1.01 sec. & 70.168 MB & $63 \cdot 10^3$ & $615 \cdot 10^3$ & 199 \\
   \hline
\textbf{32}  & N=5,T=3,Fp=0,Fs=0 & corr & \checkmark & 1.16 sec. & 70.363 MB & $69 \cdot 10^3$ & $670 \cdot 10^3$ & 199 \\
   \hline
\textbf{33}  & N=5,T=3,Fp=0,Fs=0 & relay & \checkmark & 1.61 sec. & 70.754 MB & $81 \cdot 10^3$ & $967 \cdot 10^3$ & 199 \\
   \hline
\textbf{34}  & N=5,T=3,Fp=1,Fs=0 & unforg & \checkmark & 0.05 sec. & 68.019 MB & $4 \cdot 10^3$ & $31 \cdot 10^3$ & 141 \\
   \hline
\textbf{35}  & N=5,T=3,Fp=1,Fs=0 & corr & \checkmark & 0.06 sec. & 68.019 MB & $4 \cdot 10^3$ & $40 \cdot 10^3$ & 141 \\
   \hline
\textbf{36}  & N=5,T=3,Fp=1,Fs=0 & relay & \checkmark & 0.06 sec. & 68.019 MB & $4 \cdot 10^3$ & $37 \cdot 10^3$ & 143 \\
   \hline
\textbf{37}  & N=5,T=3,Fp=1,Fs=1 & unforg & \checkmark & 0.11 sec. & 68.215 MB & $9 \cdot 10^3$ & $73 \cdot 10^3$ & 159 \\
   \hline
\textbf{38}  & N=5,T=3,Fp=1,Fs=1 & corr & \checkmark & 0.12 sec. & 68.215 MB & $10 \cdot 10^3$ & $77 \cdot 10^3$ & 159 \\
   \hline
\textbf{39}  & N=5,T=3,Fp=1,Fs=1 & relay & \checkmark & 0.17 sec. & 68.41 MB & $12 \cdot 10^3$ & $112 \cdot 10^3$ & 159 \\
   \hline
\textbf{40}  & N=5,T=3,Fp=2,Fs=0 & unforg & \checkmark & 0.01 sec. & 68.019 MB & 323 & $1 \cdot 10^3$ & 90 \\
   \hline
\textbf{41}  & N=5,T=3,Fp=2,Fs=0 & corr & \xmark & 0.01 sec. & 68.019 MB & 296 & $1 \cdot 10^3$ & 94 \\
   \hline
\textbf{42}  & N=5,T=3,Fp=2,Fs=0 & relay & \checkmark & 0.01 sec. & 68.019 MB & 322 & $1 \cdot 10^3$ & 90 \\
   \hline
\textbf{43}  & N=5,T=3,Fp=2,Fs=1 & unforg & \checkmark & 0.01 sec. & 68.019 MB & 710 & $4 \cdot 10^3$ & 107 \\
   \hline
\textbf{44}  & N=5,T=3,Fp=2,Fs=1 & corr & \checkmark & 0.01 sec. & 68.019 MB & 871 & $4 \cdot 10^3$ & 107 \\
   \hline
\textbf{45}  & N=5,T=3,Fp=2,Fs=1 & relay & \checkmark & 0.01 sec. & 68.019 MB & 763 & $4 \cdot 10^3$ & 109 \\
   \hline
\textbf{46}  & N=5,T=3,Fp=2,Fs=2 & unforg & \checkmark & 0.01 sec. & 68.019 MB & $1 \cdot 10^3$ & $7 \cdot 10^3$ & 119 \\
   \hline
\textbf{47}  & N=5,T=3,Fp=2,Fs=2 & corr & \checkmark & 0.01 sec. & 68.019 MB & $1 \cdot 10^3$ & $7 \cdot 10^3$ & 119 \\
   \hline
\textbf{48}  & N=5,T=3,Fp=2,Fs=2 & relay & \checkmark & 0.02 sec. & 68.019 MB & $1 \cdot 10^3$ & $10 \cdot 10^3$ & 119 \\
   \hline
\textbf{49}  & N=5,T=3,Fp=3,Fs=0 & unforg & \checkmark & 0.01 sec. & 68.019 MB & 35 & 135 & 48 \\
   \hline
\textbf{50}  & N=5,T=3,Fp=3,Fs=0 & corr & \xmark & 0.01 sec. & 68.019 MB & 35 & 120 & 52 \\
   \hline
\textbf{51}  & N=5,T=3,Fp=3,Fs=0 & relay & \checkmark & 0.01 sec. & 68.019 MB & 34 & 127 & 48 \\
   \hline
\rowcolor{gray}\textbf{52}  & N=5,T=3,Fp=3,Fs=1 & unforg & \checkmark & 0.01 sec. & 68.019 MB & 66 & 267 & 62 \\
   \hline
\rowcolor{gray}\textbf{53}  & N=5,T=3,Fp=3,Fs=1 & corr & \xmark & 0.01 sec. & 68.019 MB & 62 & 221 & 66 \\
   \hline
\rowcolor{gray}\textbf{54}  & N=5,T=3,Fp=3,Fs=1 & relay & \checkmark & 0.01 sec. & 68.019 MB & 62 & 235 & 62 \\
   \hline
\textbf{55}  & N=5,T=3,Fp=3,Fs=2 & unforg & \checkmark & 0.01 sec. & 68.019 MB & 107 & 447 & 73 \\
   \hline
\textbf{56}  & N=5,T=3,Fp=3,Fs=2 & corr & \checkmark & 0.01 sec. & 68.019 MB & 130 & 465 & 73 \\
   \hline
\textbf{57}  & N=5,T=3,Fp=3,Fs=2 & relay & \checkmark & 0.01 sec. & 68.019 MB & 107 & 439 & 75 \\
   \hline
\textbf{58}  & N=5,T=3,Fp=3,Fs=3 & unforg & \checkmark & 0.01 sec. & 68.019 MB & 148 & 635 & 79 \\
   \hline
\textbf{59}  & N=5,T=3,Fp=3,Fs=3 & corr & \checkmark & 0.01 sec. & 68.019 MB & 166 & 597 & 79 \\
   \hline
\textbf{60}  & N=5,T=3,Fp=3,Fs=3 & relay & \checkmark & 0.01 sec. & 68.019 MB & 163 & 727 & 79 \\
   \hline
\textbf{61}  & N=11,T=5,Fs=5,Fp=0 & unforg &  OOM & 3.45e+03 sec. & 3015.621 MB & $59 \cdot 10^6$ & 1.3010617e+09 & $1 \cdot 10^3$ \\
   \hline
\textbf{62}  & N=11,T=5,Fs=5,Fp=0 & corr &  OOM & 3.46e+03 sec. & 3015.816 MB & $59 \cdot 10^6$ & 1.3011555e+09 & $1 \cdot 10^3$ \\
   \hline
\textbf{63}  & N=11,T=5,Fs=5,Fp=0 & relay &  OOM & 2.88e+03 sec. & 3015.816 MB & $59 \cdot 10^6$ & 1.1038422e+09 & $1 \cdot 10^3$ \\
   \hline
\caption{Symm}
  \end{longtable}

%% file: t-clean-concrete.tex
\begin{longtable}{llllllllllllllll}
   \hline
\textbf{\#}  & \textbf{\scriptsize{param}} & \textbf{\scriptsize{spec}} & \textbf{\scriptsize{valid}} & \textbf{\scriptsize{SpinTime}} & \textbf{\scriptsize{SpinMemory}} & \textbf{\scriptsize{Stored}} & \textbf{\scriptsize{Transitions}} & \textbf{\scriptsize{Depth}} \\
   \hline
\textbf{1} & N=2,Tc=1,Fc=1,Fnc=0 & unforg & \checkmark & 0.01 sec. & 68.019 MB & 72 & 533 & 46 \\
   \hline
\textbf{2} & N=2,Tc=1,Fc=1,Fnc=0 & corr & \checkmark & 0.01 sec. & 68.019 MB & 131 & 863 & 50 \\
   \hline
\textbf{3} & N=2,Tc=1,Fc=1,Fnc=0 & relay & \checkmark & 0.01 sec. & 68.019 MB & 114 & $1 \cdot 10^3$ & 50 \\
   \hline
\textbf{4} & N=3,Tc=1,Fc=0,Fnc=0 & unforg & \checkmark & 0.01 sec. & 68.019 MB & 578 & $5 \cdot 10^3$ & 83 \\
   \hline
\textbf{5} & N=3,Tc=1,Fc=0,Fnc=0 & corr & \checkmark & 0.01 sec. & 68.019 MB & 759 & $7 \cdot 10^3$ & 83 \\
   \hline
\textbf{6} & N=3,Tc=1,Fc=0,Fnc=0 & relay & \checkmark & 0.01 sec. & 68.019 MB & 829 & $11 \cdot 10^3$ & 83 \\
   \hline
\textbf{7} & N=3,Tc=1,Fc=1,Fnc=0 & unforg & \checkmark & 0.01 sec. & 68.019 MB & 578 & $5 \cdot 10^3$ & 83 \\
   \hline
\textbf{8} & N=3,Tc=1,Fc=1,Fnc=0 & corr & \checkmark & 0.01 sec. & 68.019 MB & 759 & $7 \cdot 10^3$ & 83 \\
   \hline
\textbf{9} & N=3,Tc=1,Fc=1,Fnc=0 & relay & \checkmark & 0.01 sec. & 68.019 MB & 829 & $11 \cdot 10^3$ & 83 \\
   \hline
\textbf{10} & N=3,Tc=1,Fc=1,Fnc=1 & unforg & \checkmark & 0.01 sec. & 68.019 MB & 249 & $2 \cdot 10^3$ & 84 \\
   \hline
\textbf{11} & N=3,Tc=1,Fc=1,Fnc=1 & corr & \checkmark & 0.01 sec. & 68.019 MB & 424 & $4 \cdot 10^3$ & 76 \\
   \hline
\textbf{12} & N=3,Tc=1,Fc=1,Fnc=1 & relay & \checkmark & 0.01 sec. & 68.019 MB & 332 & $4 \cdot 10^3$ & 77 \\
   \hline
\textbf{13} & N=3,Tc=2,Fc=0,Fnc=0 & unforg & \checkmark & 0.01 sec. & 68.019 MB & 668 & $7 \cdot 10^3$ & 77 \\
   \hline
\textbf{14} & N=3,Tc=2,Fc=0,Fnc=0 & corr & \checkmark & 0.02 sec. & 68.019 MB & 892 & $8 \cdot 10^3$ & 81 \\
   \hline
\textbf{15} & N=3,Tc=2,Fc=0,Fnc=0 & relay & \checkmark & 0.02 sec. & 68.019 MB & $1 \cdot 10^3$ & $17 \cdot 10^3$ & 81 \\
   \hline
\textbf{16} & N=3,Tc=2,Fc=1,Fnc=0 & unforg & \checkmark & 0.01 sec. & 68.019 MB & 668 & $7 \cdot 10^3$ & 77 \\
   \hline
\textbf{17} & N=3,Tc=2,Fc=1,Fnc=0 & corr & \checkmark & 0.01 sec. & 68.019 MB & 892 & $8 \cdot 10^3$ & 81 \\
   \hline
\textbf{18} & N=3,Tc=2,Fc=1,Fnc=0 & relay & \checkmark & 0.02 sec. & 68.019 MB & $1 \cdot 10^3$ & $17 \cdot 10^3$ & 81 \\
   \hline
\textbf{19} & N=3,Tc=2,Fc=1,Fnc=1 & unforg & \checkmark & 0.01 sec. & 68.019 MB & 279 & $2 \cdot 10^3$ & 77 \\
   \hline
\textbf{20} & N=3,Tc=2,Fc=1,Fnc=1 & corr & \checkmark & 0.01 sec. & 68.019 MB & 425 & $4 \cdot 10^3$ & 78 \\
   \hline
\textbf{21} & N=3,Tc=2,Fc=1,Fnc=1 & relay & \checkmark & 0.01 sec. & 68.019 MB & 475 & $6 \cdot 10^3$ & 78 \\
   \hline
\rowcolor{gray}\textbf{22} & N=3,Tc=2,Fc=2,Fnc=0 & unforg & \checkmark & 0.01 sec. & 68.019 MB & 668 & $7 \cdot 10^3$ & 77 \\
   \hline
\rowcolor{gray}\textbf{23} & N=3,Tc=2,Fc=2,Fnc=0 & corr & \checkmark & 0.01 sec. & 68.019 MB & 892 & $8 \cdot 10^3$ & 81 \\
   \hline
\rowcolor{gray}\textbf{24} & N=3,Tc=2,Fc=2,Fnc=0 & relay & \checkmark & 0.02 sec. & 68.019 MB & $1 \cdot 10^3$ & $17 \cdot 10^3$ & 81 \\
   \hline
\textbf{25} & N=3,Tc=2,Fc=2,Fnc=1 & unforg & \checkmark & 0.01 sec. & 68.019 MB & 279 & $2 \cdot 10^3$ & 77 \\
   \hline
\textbf{26} & N=3,Tc=2,Fc=2,Fnc=1 & corr & \checkmark & 0.01 sec. & 68.019 MB & 425 & $4 \cdot 10^3$ & 78 \\
   \hline
\textbf{27} & N=3,Tc=2,Fc=2,Fnc=1 & relay & \checkmark & 0.01 sec. & 68.019 MB & 475 & $6 \cdot 10^3$ & 78 \\
   \hline
\textbf{28} & N=3,Tc=2,Fc=2,Fnc=2 & unforg & \checkmark & 0.01 sec. & 68.019 MB & 133 & $1 \cdot 10^3$ & 72 \\
   \hline
\textbf{29} & N=3,Tc=2,Fc=2,Fnc=2 & corr & \checkmark & 0.01 sec. & 68.019 MB & 216 & $2 \cdot 10^3$ & 71 \\
   \hline
\textbf{30} & N=3,Tc=2,Fc=2,Fnc=2 & relay & \checkmark & 0.01 sec. & 68.019 MB & 198 & $2 \cdot 10^3$ & 72 \\
   \hline
\textbf{31} & N=3,Tc=3,Fc=0,Fnc=0 & unforg & \xmark & 0.03 sec. & 68.019 MB & 561 & $6 \cdot 10^3$ & 78 \\
   \hline
\textbf{32} & N=3,Tc=3,Fc=0,Fnc=0 & corr & \checkmark & 0.01 sec. & 68.019 MB & $1 \cdot 10^3$ & $12 \cdot 10^3$ & 76 \\
   \hline
\textbf{33} & N=3,Tc=3,Fc=0,Fnc=0 & relay & \checkmark & 0.04 sec. & 68.019 MB & $1 \cdot 10^3$ & $29 \cdot 10^3$ & 76 \\
   \hline
\textbf{34} & N=3,Tc=3,Fc=1,Fnc=0 & unforg & \xmark & 0.01 sec. & 68.019 MB & 561 & $6 \cdot 10^3$ & 78 \\
   \hline
\textbf{35} & N=3,Tc=3,Fc=1,Fnc=0 & corr & \checkmark & 0.02 sec. & 68.019 MB & $1 \cdot 10^3$ & $12 \cdot 10^3$ & 76 \\
   \hline
\textbf{36} & N=3,Tc=3,Fc=1,Fnc=0 & relay & \checkmark & 0.04 sec. & 68.019 MB & $1 \cdot 10^3$ & $29 \cdot 10^3$ & 76 \\
   \hline
\textbf{37} & N=3,Tc=3,Fc=1,Fnc=1 & unforg & \checkmark & 0.01 sec. & 68.019 MB & 529 & $5 \cdot 10^3$ & 69 \\
   \hline
\textbf{38} & N=3,Tc=3,Fc=1,Fnc=1 & corr & \checkmark & 0.01 sec. & 68.019 MB & 768 & $7 \cdot 10^3$ & 73 \\
   \hline
\textbf{39} & N=3,Tc=3,Fc=1,Fnc=1 & relay & \xmark & 0.01 sec. & 68.019 MB & 92 & $1 \cdot 10^3$ & 54 \\
   \hline
\textbf{40} & N=3,Tc=3,Fc=2,Fnc=0 & unforg & \xmark & 0.01 sec. & 68.019 MB & 561 & $6 \cdot 10^3$ & 78 \\
   \hline
\textbf{41} & N=3,Tc=3,Fc=2,Fnc=0 & corr & \checkmark & 0.02 sec. & 68.019 MB & $1 \cdot 10^3$ & $12 \cdot 10^3$ & 76 \\
   \hline
\textbf{42} & N=3,Tc=3,Fc=2,Fnc=0 & relay & \checkmark & 0.04 sec. & 68.019 MB & $1 \cdot 10^3$ & $29 \cdot 10^3$ & 76 \\
   \hline
\textbf{43} & N=3,Tc=3,Fc=2,Fnc=1 & unforg & \checkmark & 0.01 sec. & 68.019 MB & 529 & $5 \cdot 10^3$ & 69 \\
   \hline
\textbf{44} & N=3,Tc=3,Fc=2,Fnc=1 & corr & \checkmark & 0.02 sec. & 68.019 MB & 768 & $7 \cdot 10^3$ & 73 \\
   \hline
\textbf{45} & N=3,Tc=3,Fc=2,Fnc=1 & relay & \xmark & 0.01 sec. & 68.019 MB & 92 & $1 \cdot 10^3$ & 54 \\
   \hline
\textbf{46} & N=3,Tc=3,Fc=2,Fnc=2 & unforg & \checkmark & 0.01 sec. & 68.019 MB & 275 & $2 \cdot 10^3$ & 66 \\
   \hline
\textbf{47} & N=3,Tc=3,Fc=2,Fnc=2 & corr & \checkmark & 0.01 sec. & 68.019 MB & 446 & $5 \cdot 10^3$ & 70 \\
   \hline
\textbf{48} & N=3,Tc=3,Fc=2,Fnc=2 & relay & \xmark & 0.01 sec. & 68.019 MB & 12 & 86 & 31 \\
   \hline
\textbf{49} & N=3,Tc=3,Fc=3,Fnc=0 & unforg & \xmark & 0.01 sec. & 68.019 MB & 561 & $6 \cdot 10^3$ & 78 \\
   \hline
\textbf{50} & N=3,Tc=3,Fc=3,Fnc=0 & corr & \checkmark & 0.02 sec. & 68.019 MB & $1 \cdot 10^3$ & $12 \cdot 10^3$ & 76 \\
   \hline
\textbf{51} & N=3,Tc=3,Fc=3,Fnc=0 & relay & \checkmark & 0.03 sec. & 68.019 MB & $1 \cdot 10^3$ & $29 \cdot 10^3$ & 76 \\
   \hline
\textbf{52} & N=3,Tc=3,Fc=3,Fnc=1 & unforg & \checkmark & 0.01 sec. & 68.019 MB & 529 & $5 \cdot 10^3$ & 69 \\
   \hline
\textbf{53} & N=3,Tc=3,Fc=3,Fnc=1 & corr & \checkmark & 0.01 sec. & 68.019 MB & 768 & $7 \cdot 10^3$ & 73 \\
   \hline
\textbf{54} & N=3,Tc=3,Fc=3,Fnc=1 & relay & \xmark & 0.01 sec. & 68.019 MB & 92 & $1 \cdot 10^3$ & 54 \\
   \hline
\textbf{55} & N=3,Tc=3,Fc=3,Fnc=2 & unforg & \checkmark & 0.01 sec. & 68.019 MB & 275 & $2 \cdot 10^3$ & 66 \\
   \hline
\textbf{56} & N=3,Tc=3,Fc=3,Fnc=2 & corr & \checkmark & 0.01 sec. & 68.019 MB & 446 & $5 \cdot 10^3$ & 70 \\
   \hline
\textbf{57} & N=3,Tc=3,Fc=3,Fnc=2 & relay & \xmark & 0.01 sec. & 68.019 MB & 12 & 86 & 31 \\
   \hline
\textbf{58} & N=3,Tc=3,Fc=3,Fnc=3 & unforg & \checkmark & 0.01 sec. & 68.019 MB & 238 & $2 \cdot 10^3$ & 49 \\
   \hline
\textbf{59} & N=3,Tc=3,Fc=3,Fnc=3 & corr & \xmark & 0.01 sec. & 68.019 MB & 249 & $3 \cdot 10^3$ & 53 \\
   \hline
\textbf{60} & N=3,Tc=3,Fc=3,Fnc=3 & relay & \xmark & 0.01 sec. & 68.019 MB & 12 & 86 & 31 \\
   \hline
\textbf{61} & N=11,Tc=10,Fc=10,Fnc=0 & unforg &  OOM & 6.75e+03 sec. & 3015.621 MB & $59 \cdot 10^6$ & 2.6021236e+09 & 757 \\
   \hline
\textbf{62} & N=11,Tc=10,Fc=10,Fnc=0 & corr &  OOM & 6.67e+03 sec. & 3015.621 MB & $59 \cdot 10^6$ & 2.6021237e+09 & 761 \\
   \hline
\textbf{63} & N=11,Tc=10,Fc=10,Fnc=0 & relay &  OOM & 6.72e+03 sec. & 3015.621 MB & $59 \cdot 10^6$ & 2.6055872e+09 & 761 \\
   \hline
\caption{Clean} 
\end{longtable}

%% file: t-byz-concrete.tex
\begin{table*}
 \begin{center}
  \begin{tabular}{llllllllllllllll}
   \hline
\textbf{\#}  & \textbf{\scriptsize{param}} & \textbf{\scriptsize{spec}} & \textbf{\scriptsize{valid}} & \textbf{\scriptsize{SpinTime}} & \textbf{\scriptsize{SpinMemory}} & \textbf{\scriptsize{Stored}} & \textbf{\scriptsize{Transitions}} & \textbf{\scriptsize{Depth}} \\
   \hline
\textbf{1}  & N=4,T=1,F=1 & unforg & \checkmark & 0.01 sec. & 68.019 MB & 533 & $3 \cdot 10^3$ & 82 \\
   \hline
\textbf{2}  & N=4,T=1,F=1 & corr & \checkmark & 0.01 sec. & 68.019 MB & 639 & $3 \cdot 10^3$ & 82 \\
   \hline
\textbf{3}  & N=4,T=1,F=1 & relay & \checkmark & 0.01 sec. & 68.019 MB & 706 & $5 \cdot 10^3$ & 82 \\
   \hline
\textbf{4}  & N=7,T=1,F=0 & unforg & \checkmark & 278 sec. & 458.254 MB & $10 \cdot 10^6$ & 1.4322797e+08 & 319 \\
   \hline
\textbf{5}  & N=7,T=1,F=0 & corr & \checkmark & 325 sec. & 513.332 MB & $11 \cdot 10^6$ & 1.6546415e+08 & 319 \\
   \hline
\textbf{6}  & N=7,T=1,F=0 & relay & \checkmark & 485 sec. & 603.371 MB & $14 \cdot 10^6$ & 2.4976492e+08 & 319 \\
   \hline
\textbf{7}  & N=7,T=1,F=1 & unforg & \checkmark & 26 sec. & 114.504 MB & $1 \cdot 10^6$ & $14 \cdot 10^6$ & 268 \\
   \hline
\textbf{8}  & N=7,T=1,F=1 & corr & \checkmark & 31.2 sec. & 122.121 MB & $1 \cdot 10^6$ & $17 \cdot 10^6$ & 268 \\
   \hline
\textbf{9}  & N=7,T=1,F=1 & relay & \checkmark & 44.7 sec. & 130.91 MB & $1 \cdot 10^6$ & $25 \cdot 10^6$ & 268 \\
   \hline
\textbf{10}  & N=7,T=1,F=2 & unforg & \xmark & 1.21 sec. & 70.558 MB & $74 \cdot 10^3$ & $748 \cdot 10^3$ & 241 \\
   \hline
\textbf{11}  & N=7,T=1,F=2 & corr & \xmark & 2.08 sec. & 72.316 MB & $127 \cdot 10^3$ & $1 \cdot 10^6$ & 225 \\
   \hline
\textbf{12}  & N=7,T=1,F=2 & relay & \xmark & 0.01 sec. & 68.019 MB & 42 & 196 & 222 \\
   \hline
\textbf{13}  & N=7,T=1,F=3 & unforg & \xmark & 0.09 sec. & 68.215 MB & $7 \cdot 10^3$ & $63 \cdot 10^3$ & 201 \\
   \hline
\textbf{14}  & N=7,T=1,F=3 & corr & \xmark & 0.16 sec. & 68.41 MB & $13 \cdot 10^3$ & $107 \cdot 10^3$ & 181 \\
   \hline
\textbf{15}  & N=7,T=1,F=3 & relay & \xmark & 0.01 sec. & 68.019 MB & 35 & 127 & 182 \\
   \hline
\textbf{16}  & N=7,T=2,F=0 & unforg & \checkmark & 416 sec. & 643.215 MB & $15 \cdot 10^6$ & 2.111452e+08 & 325 \\
   \hline
\textbf{17}  & N=7,T=2,F=0 & corr & \checkmark & 435 sec. & 665.48 MB & $15 \cdot 10^6$ & 2.1891381e+08 & 325 \\
   \hline
\textbf{18}  & N=7,T=2,F=0 & relay & \checkmark & 859 sec. & 949.66 MB & $23 \cdot 10^6$ & 4.3596991e+08 & 325 \\
   \hline
\textbf{19}  & N=7,T=2,F=1 & unforg & \checkmark & 38.1 sec. & 135.597 MB & $1 \cdot 10^6$ & $21 \cdot 10^6$ & 273 \\
   \hline
\textbf{20}  & N=7,T=2,F=1 & corr & \checkmark & 40.3 sec. & 139.113 MB & $1 \cdot 10^6$ & $22 \cdot 10^6$ & 273 \\
   \hline
\textbf{21}  & N=7,T=2,F=1 & relay & \checkmark & 77.9 sec. & 170.363 MB & $2 \cdot 10^6$ & $43 \cdot 10^6$ & 273 \\
   \hline
\rowcolor{gray}\textbf{22}  & N=7,T=2,F=2 & unforg & \checkmark & 3.13 sec. & 74.66 MB & $193 \cdot 10^3$ & $1 \cdot 10^6$ & 229 \\
   \hline
\rowcolor{gray}\textbf{23}  & N=7,T=2,F=2 & corr & \checkmark & 3.43 sec. & 75.051 MB & $207 \cdot 10^3$ & $2 \cdot 10^6$ & 229 \\
   \hline
\rowcolor{gray}\textbf{24}  & N=7,T=2,F=2 & relay & \checkmark & 6.3 sec. & 77.98 MB & $290 \cdot 10^3$ & $3 \cdot 10^6$ & 229 \\
   \hline
\textbf{25}  & N=7,T=2,F=3 & unforg & \xmark & 0.11 sec. & 68.215 MB & $7 \cdot 10^3$ & $60 \cdot 10^3$ & 205 \\
   \hline
\textbf{26}  & N=7,T=2,F=3 & corr & \xmark & 0.21 sec. & 68.605 MB & $18 \cdot 10^3$ & $143 \cdot 10^3$ & 185 \\
   \hline
\textbf{27}  & N=7,T=2,F=3 & relay & \xmark & 0.01 sec. & 68.019 MB & 33 & 119 & 176 \\
   \hline
\textbf{28}  & N=7,T=3,F=0 & unforg & \checkmark & 596 sec. & 876.418 MB & $21 \cdot 10^6$ & 2.967337e+08 & 331 \\
   \hline
\textbf{29}  & N=7,T=3,F=0 & corr & \checkmark & 604 sec. & 883.449 MB & $21 \cdot 10^6$ & 2.9891686e+08 & 331 \\
   \hline
\textbf{30}  & N=7,T=3,F=0 & relay & \checkmark & 1.43e+03 sec. & 1678.902 MB & $35 \cdot 10^6$ & 7.09111e+08 & 331 \\
   \hline
\textbf{31}  & N=7,T=3,F=1 & unforg & \checkmark & 55.5 sec. & 162.551 MB & $2 \cdot 10^6$ & $29 \cdot 10^6$ & 278 \\
   \hline
\textbf{32}  & N=7,T=3,F=1 & corr & \checkmark & 56 sec. & 163.722 MB & $2 \cdot 10^6$ & $30 \cdot 10^6$ & 278 \\
   \hline
\textbf{33}  & N=7,T=3,F=1 & relay & \xmark & 0.83 sec. & 68.996 MB & $29 \cdot 10^3$ & $461 \cdot 10^3$ & 278 \\
   \hline
\rowcolor{gray}\textbf{34}  & N=7,T=3,F=2 & unforg & \checkmark & 4.38 sec. & 77.004 MB & $265 \cdot 10^3$ & $2 \cdot 10^6$ & 233 \\
   \hline
\rowcolor{gray}\textbf{35}  & N=7,T=3,F=2 & corr & \checkmark & 4.5 sec. & 77.199 MB & $271 \cdot 10^3$ & $2 \cdot 10^6$ & 233 \\
   \hline
\rowcolor{gray}\textbf{36}  & N=7,T=3,F=2 & relay & \xmark & 0.02 sec. & 68.019 MB & $1 \cdot 10^3$ & $13 \cdot 10^3$ & 210 \\
   \hline
\textbf{37}  & N=7,T=3,F=3 & unforg & \checkmark & 0.32 sec. & 68.801 MB & $25 \cdot 10^3$ & $209 \cdot 10^3$ & 188 \\
   \hline
\textbf{38}  & N=7,T=3,F=3 & corr & \checkmark & 0.33 sec. & 68.801 MB & $26 \cdot 10^3$ & $215 \cdot 10^3$ & 188 \\
   \hline
\textbf{39}  & N=7,T=3,F=3 & relay & \xmark & 0.01 sec. & 68.019 MB & 100 & 528 & 165 \\
   \hline
\textbf{40}  & N=10,T=3,F=3 & unforg & OOM & 2.02e+03 sec. & 3015.621 MB & $70 \cdot 10^6$ & 9.9379608e+08 & 452 \\
   \hline
\textbf{41}  & N=10,T=3,F=3 & corr &  OOM & 2.12e+03 sec. & 3015.816 MB & $70 \cdot 10^6$ & 9.9381042e+08 & 452 \\
   \hline
\textbf{42}  & N=10,T=3,F=3 & relay &  OOM & 2.97e+03 sec. & 3015.816 MB & $70 \cdot 10^6$ & 1.4335274e+09 & 452 \\
   \hline
  \end{tabular}
 \end{center}
\caption{Byz}
\end{table*}

%% file: t-omit-concrete.tex
\begin{table*}
 \begin{center}
  \begin{tabular}{llllllllllllllll}
   \hline
\textbf{\#}  & \textbf{\scriptsize{param}} & \textbf{\scriptsize{spec}} & \textbf{\scriptsize{valid}} & \textbf{\scriptsize{SpinTime}} & \textbf{\scriptsize{SpinMemory}} & \textbf{\scriptsize{Stored}} & \textbf{\scriptsize{Transitions}} & \textbf{\scriptsize{Depth}} \\
   \hline
\textbf{1}  & N=3,To=1,Fo=1 & unforg & \checkmark & 0.01 sec. & 68.019 MB & 440 & $4 \cdot 10^3$ & 77 \\
   \hline
\textbf{2}  & N=3,To=1,Fo=1 & corr & \checkmark & 0.01 sec. & 68.019 MB & 691 & $7 \cdot 10^3$ & 85 \\
   \hline
\textbf{3}  & N=3,To=1,Fo=1 & relay & \checkmark & 0.01 sec. & 68.019 MB & 731 & $10 \cdot 10^3$ & 85 \\
   \hline
\textbf{4}  & N=5,To=1,Fo=0 & unforg & \checkmark & 1.46 sec. & 69.582 MB & $51 \cdot 10^3$ & $878 \cdot 10^3$ & 175 \\
   \hline
\textbf{5}  & N=5,To=1,Fo=0 & corr & \checkmark & 1.41 sec. & 69.777 MB & $52 \cdot 10^3$ & $891 \cdot 10^3$ & 179 \\
   \hline
\textbf{6}  & N=5,To=1,Fo=0 & relay & \checkmark & 3.85 sec. & 71.144 MB & $96 \cdot 10^3$ & $2 \cdot 10^6$ & 179 \\
   \hline
\textbf{7}  & N=5,To=1,Fo=1 & unforg & \checkmark & 1.38 sec. & 69.582 MB & $51 \cdot 10^3$ & $878 \cdot 10^3$ & 175 \\
   \hline
\textbf{8}  & N=5,To=1,Fo=1 & corr & \checkmark & 1.42 sec. & 69.777 MB & $53 \cdot 10^3$ & $909 \cdot 10^3$ & 183 \\
   \hline
\textbf{9}  & N=5,To=1,Fo=1 & relay & \checkmark & 3.86 sec. & 71.34 MB & $97 \cdot 10^3$ & $2 \cdot 10^6$ & 183 \\
   \hline
\textbf{10}  & N=5,To=1,Fo=2 & unforg & \checkmark & 1.38 sec. & 69.582 MB & $51 \cdot 10^3$ & $878 \cdot 10^3$ & 175 \\
   \hline
\textbf{11}  & N=5,To=1,Fo=2 & corr & \checkmark & 1.54 sec. & 69.777 MB & $56 \cdot 10^3$ & $979 \cdot 10^3$ & 183 \\
   \hline
\textbf{12}  & N=5,To=1,Fo=2 & relay & \xmark & 0.01 sec. & 68.019 MB & 15 & 131 & 42 \\
   \hline
\textbf{13}  & N=5,To=1,Fo=3 & unforg & \checkmark & 1.39 sec. & 69.582 MB & $51 \cdot 10^3$ & $878 \cdot 10^3$ & 175 \\
   \hline
\textbf{14}  & N=5,To=1,Fo=3 & corr & \checkmark & 1.88 sec. & 70.168 MB & $62 \cdot 10^3$ & $1 \cdot 10^6$ & 183 \\
   \hline
\textbf{15}  & N=5,To=1,Fo=3 & relay & \xmark & 0.01 sec. & 68.019 MB & 15 & 131 & 42 \\
   \hline
\textbf{16}  & N=5,To=2,Fo=0 & unforg & \checkmark & 1.37 sec. & 69.582 MB & $51 \cdot 10^3$ & $878 \cdot 10^3$ & 175 \\
   \hline
\textbf{17}  & N=5,To=2,Fo=0 & corr & \checkmark & 1.49 sec. & 69.972 MB & $57 \cdot 10^3$ & $945 \cdot 10^3$ & 179 \\
   \hline
\textbf{18}  & N=5,To=2,Fo=0 & relay & \checkmark & 3.56 sec. & 70.949 MB & $88 \cdot 10^3$ & $2 \cdot 10^6$ & 179 \\
   \hline
\textbf{19}  & N=5,To=2,Fo=1 & unforg & \checkmark & 1.38 sec. & 69.582 MB & $51 \cdot 10^3$ & $878 \cdot 10^3$ & 175 \\
   \hline
\textbf{20}  & N=5,To=2,Fo=1 & corr & \checkmark & 1.51 sec. & 69.972 MB & $58 \cdot 10^3$ & $963 \cdot 10^3$ & 183 \\
   \hline
\textbf{21}  & N=5,To=2,Fo=1 & relay & \checkmark & 3.53 sec. & 70.949 MB & $89 \cdot 10^3$ & $2 \cdot 10^6$ & 183 \\
   \hline
\rowcolor{gray}\textbf{22}  & N=5,To=2,Fo=2 & unforg & \checkmark & 1.43 sec. & 69.582 MB & $51 \cdot 10^3$ & $878 \cdot 10^3$ & 175 \\
   \hline
\rowcolor{gray}\textbf{23}  & N=5,To=2,Fo=2 & corr & \checkmark & 1.64 sec. & 69.972 MB & $60 \cdot 10^3$ & $1 \cdot 10^6$ & 183 \\
   \hline
\rowcolor{gray}\textbf{24}  & N=5,To=2,Fo=2 & relay & \checkmark & 3.69 sec. & 71.144 MB & $92 \cdot 10^3$ & $2 \cdot 10^6$ & 183 \\
   \hline
\rowcolor{gray}\textbf{25}  & N=5,To=2,Fo=3 & unforg & \checkmark & 1.39 sec. & 69.582 MB & $51 \cdot 10^3$ & $878 \cdot 10^3$ & 175 \\
   \hline
\rowcolor{gray}\textbf{26}  & N=5,To=2,Fo=3 & corr & \xmark & 1.63 sec. & 69.777 MB & $53 \cdot 10^3$ & $1 \cdot 10^6$ & 183 \\
   \hline
\rowcolor{gray}\textbf{27}  & N=5,To=2,Fo=3 & relay & \xmark & 0.01 sec. & 68.019 MB & 17 & 135 & 53 \\
   \hline
\textbf{28}  & N=5,To=3,Fo=0 & unforg & \checkmark & 1.41 sec. & 69.582 MB & $51 \cdot 10^3$ & $878 \cdot 10^3$ & 175 \\
   \hline
\textbf{29}  & N=5,To=3,Fo=0 & corr & \checkmark & 1.8 sec. & 70.363 MB & $69 \cdot 10^3$ & $1 \cdot 10^6$ & 179 \\
   \hline
\textbf{30}  & N=5,To=3,Fo=0 & relay & \checkmark & 2.9 sec. & 70.558 MB & $75 \cdot 10^3$ & $1 \cdot 10^6$ & 179 \\
   \hline
\textbf{31}  & N=5,To=3,Fo=1 & unforg & \checkmark & 1.39 sec. & 69.582 MB & $51 \cdot 10^3$ & $878 \cdot 10^3$ & 175 \\
   \hline
\textbf{32}  & N=5,To=3,Fo=1 & corr & \checkmark & 1.83 sec. & 70.363 MB & $70 \cdot 10^3$ & $1 \cdot 10^6$ & 183 \\
   \hline
\textbf{33}  & N=5,To=3,Fo=1 & relay & \checkmark & 2.89 sec. & 70.558 MB & $76 \cdot 10^3$ & $1 \cdot 10^6$ & 183 \\
   \hline
\textbf{34}  & N=5,To=3,Fo=2 & unforg & \checkmark & 1.4 sec. & 69.582 MB & $51 \cdot 10^3$ & $878 \cdot 10^3$ & 175 \\
   \hline
\textbf{35}  & N=5,To=3,Fo=2 & corr & \xmark & 1.37 sec. & 69.582 MB & $47 \cdot 10^3$ & $851 \cdot 10^3$ & 183 \\
   \hline
\textbf{36}  & N=5,To=3,Fo=2 & relay & \xmark & 0.01 sec. & 68.019 MB & 38 & 257 & 171 \\
   \hline
\textbf{37}  & N=5,To=3,Fo=3 & unforg & \checkmark & 1.39 sec. & 69.582 MB & $51 \cdot 10^3$ & $878 \cdot 10^3$ & 175 \\
   \hline
\textbf{38}  & N=5,To=3,Fo=3 & corr & \xmark & 1.65 sec. & 69.777 MB & $53 \cdot 10^3$ & $1 \cdot 10^6$ & 183 \\
   \hline
\textbf{39}  & N=5,To=3,Fo=3 & relay & \xmark & 0.01 sec. & 68.019 MB & 38 & 257 & 171 \\
   \hline
\textbf{40} & N=11,To=5,Fo=5 & unforg & OOM & 6.97e+03 sec. & 2757.347 MB & $59 \cdot 10^6$ & 2.60E+009 & 757 \\
   \hline
\textbf{41} & N=11,To=5,Fo=5 & corr & OOM & 7.25e+03 sec. & 3015.621 MB & $59 \cdot 10^6$ & 2.7891908e+09 & 765 \\
   \hline
\textbf{42} & N=11,To=5,Fo=5 & relay & OOM & 9.82e+03 sec. & 3015.621 MB & $59 \cdot 10^6$ & 3.7808961e+09 & 765 \\
   \hline
  \end{tabular}
 \end{center}
\caption{Omit}
\end{table*}